\documentclass[floatfix,float,aps,prb,amsmath,twocolumn,amssymb,titlepage,superscriptaddress]{revtex4-1}
\usepackage{amsthm,amsmath,amsfonts,dsfont} 
\usepackage{amsmath}    
\usepackage{graphicx}   
\usepackage{verbatim}   
\usepackage{subfigure}
\usepackage{hyperref} 
\usepackage[usenames, dvipsnames]{xcolor}
\usepackage{MnSymbol}
\usepackage{tikz}
\usepackage{enumitem}
\usetikzlibrary{plotmarks,arrows.meta,calc,decorations.markings,math,arrows.meta,decorations.pathreplacing,decorations.pathmorphing,patterns}
\usepackage[percent]{overpic}
\def\be{\begin{equation}}
\def\ee{\end{equation}}

\usepackage{braket}

\newcommand{\subfig}[2]{%
    {#1} \vtop{
  \vskip0pt
  \hbox{#2}
}}

\begin{document}


\title{Kibble-Zurek mechanism and errors of gapped quantum phases}

\author{Amit Jamadagni}
\email{amit.gangapuram@lrz.de}
\affiliation{Leibniz Supercomputing Centre (LRZ), Garching, 85748, Germany}
\affiliation{Laboratory for Theoretical and Computational Physics,
Paul Scherrer Institute (PSI), 5232 Villigen, Switzerland}
\affiliation{ETH Z\"urich-PSI Quantum Computing Hub, Paul Scherrer Institut, 5232 Villigen, Switzerland}
\author{Javad Kazemi}
\email{seyedjavadkazemi@gmail.com}
\affiliation{Parity Quantum Computing Germany GmbH,
Schauenburgerstra{\ss}e 6, 
20095 Hamburg, Germany}
\author{Arpan Bhattacharyya}
\email{abhattacharyya@iitgn.ac.in}
\affiliation{Indian Institute of Technology, Gandhinagar, Gujarat 382355, India}

\begin{abstract}
Kibble-Zurek mechanism relates the domain of non-equilibrium dynamics with the critical properties at 
equilibrium. It establishes a power law connection between non-equilibrium defects quenched through a continuous
phase transition and the quench rate via the scaling exponent. We present a novel numerical scheme to
estimate the scaling exponent wherein the notion of defects is mapped to errors, previously introduced 
to quantify a variety of gapped quantum phases. To demonstrate the versatility of our method we conduct numerical experiments across a broad spectrum of spin-half models hosting local and symmetry protected topological order. 
Furthermore, an implementation of the quench dynamics featuring a topological phase transition
on a digital quantum computer is proposed to quantify the associated criticality.
\end{abstract}

\maketitle

\section{Introduction}
The exploration of quantum phases at absolute zero and transitions among them is a cornerstone of condensed matter 
physics. The classification of quantum phases is an active area of research, specially as the investigation extends beyond the principles of Landau symmetry breaking. Essentially, there is a comprehensive understanding of quantum phases characterized by local order, yet comprehending phases that go beyond local order poses significant challenges. The study of associated Quantum Phase Transitions (QPT) provides critical insight into the universal behavior linked to the divergence in the correlation length. This makes it possible to categorize phases in different universality classes being identified by a set of critical exponents.

Kibble-Zurek Mechanism (KZM) provides critical insight into the dynamics of a system driven through a continuous 
phase transition~\cite{Kibble:1976sj,Kibble:1980mv,Zurek:1985qw,Zurek:1993ek}. Introduced in the context of 
cosmological phase transitions, the importance of the KZM lies in its ability to elucidate the emergence of 
defects during rapid phase transitions \cite{Kibble:1976sj,Kibble:1980mv,Zurek:1985qw,Zurek:1993ek,Campo2014} 
while establishing a relationship between non-equilibrium dynamics and the critical exponents associated with 
phase transitions. Lately, this mechanism has been extended to encompass quantum many-body systems~\cite{Ebadi:2020ldi,Zoller2005,Dziarmaga2005DynamicsOA,PhysRevA.75.052321,Dziarmaga2009DynamicsOA,PhysRevA.83.062104,PhysRevA.100.032115,PhysRevLett.109.015701,Gong2015SimulatingTK,PhysRevB.86.064304,PhysRevLett.123.130603,PhysRevB.93.075134,Schmitt2021QuantumPT,Dupont2022}, and referred to as Quantum Kibble-Zurek Mechansim (QKZM).
QKZM describes the quantitative behaviour of the defects in situations where the rate of change of the Hamiltonian 
is faster than the inverse of the spectral gap of the underlying system. The quantity of defects evolving through 
a QPT can be measured by employing the critical properties linked to it. Specifically, the defect density follows 
a power-law relationship with the quench rate, introducing a parameter known as the scaling exponent $\mu$. 

Recent advancements in the development of several quantum architectures has propelled significant 
interest in exploring various quantum many-body phenomena~\cite{Smith2019,deLeseleuc2019,Brydges_2019,Bhattacharyya:2021ypq,2023PhRvX..13b1042W}.
In particular, QKZM has been validated on different quantum hardware platforms by estimating the 
critical properties, especially in Ising-like models~\cite{Keesling_2019,King_2021,Dupont2022,King_2022,Bando_2020,Du_2023}. 
Motivated by the recent 
progress, in this work, we introduce a novel numerical scheme to obtain the scaling exponent in QPTs that are 
in principle accessible on a quantum device. We emphasize that the introduced method is not limited
to the scope of the Landau symmetry breaking principle, and can be applied to a broader range of QPTs
involving phases characterized by non-local order. To this extent, we further present strategies that 
enable numerical and experimental observation of QKZM involving symmetry protected topological phases. 
This in turn can be used as a scheme to profile and benchmark the performance of quantum computing devices.

We structure the rest of paper as follows: in Sec.~\ref{sec:qkzm}, we briefly review a quench protocol that 
realizes the QKZM, followed by introducing expectation value based strategies to determine the defect
densities. In addition, we describe various components involved in estimating the scaling exponent in 
the thermodynamic limit. In Sec.~\ref{sec:local_order}, we start our numerical investigations by studying the 
transverse field Ising model that exhibits local order. In Sec.~\ref{sec:spt_order}, we extend the analysis to 
study phase transitions involving symmetry protected topological phases by considering
various model Hamiltonians. Further, in Sec.~\ref{sec:cric_qc}, we present an experimental prototype that 
allows the estimation of the scaling exponent associated with a topological phase transition on a 
digital quantum computer. Towards the end, in Sec.~\ref{sec:summary}, we summarize the main 
results while outlining some future directions that can be explored using the computational strategies and protocols introduced. 

\section{Quench protocol, defect density and critical exponent \label{sec:qkzm}}

We start by presenting the quench protocol as in the context of QKZM while introducing methods to compute defect densities and to estimate the corresponding scaling exponent.

\subsection{Revisiting the QKZM}
The QKZM provides a theoretical framework for understanding the behavior of a system undergoing a continuous QPT when it is driven out of equilibrium by a rapid change in a control parameter. The mechanism relies on a quench dynamics that can be generated by a time-dependent Hamiltonian, $H(t)$, connecting point $I$ to $F$,
\begin{equation}\label{eq1}
 H(t) = H_{F} + g(t)H_{I},
\end{equation}
with $g(t) = -t/\tau_{Q}$ denotes a linear quench with the rate $\tau_{Q}$ in the time interval $t\in(-\infty, 0]$~\cite{Campo2014}. Having the system prepared in the groundstate of the Hamiltonian, 
$H_{I}$, we evolve the system under the total Hamiltonian mentioned in  Eq.~\ref{eq1}. Assuming that there exists a second-order QPT at some critical strength $g(t_c)$, the correlation length $\xi$ diverges at criticality as $\xi\propto |g(t)|^{-\nu}$ with $\nu$ being the correlation-length critical exponent. This is associated with the closing of the energy gap, or the divergence of the relaxation time as $\tau\simeq\Delta E^{-1} \propto |g(t)|^{-z\nu}$ where $z$ is the dynamical critical exponent, see for example Ref.~\onlinecite{Sachdev_2011, Dziarmaga_2005} on procedures to estimate the critical exponents. The QKZM establishes a relation between the the size of the non-adiabatic domains and the critical exponents expressed as
\begin{equation}
\xi \propto \tau_{Q}^{\frac{z}{1+z\nu}}.
\end{equation}
Consequently the defect density in a $D$ dimensional system exhibits a power-law scaling as
$\eta \propto \tau_{Q}^{-\mu}$ where $\mu=Dz/(1+z\nu)$ is the QKZM scaling exponent determined by the universality class of the underlying QPT
~\cite{Zoller2005, Dziarmaga_2005, Polkovnikov_2005}. In this work, we especially focus on one-dimensional 
systems, however our method can be extended to generic gapped phases, see App.~\ref{app:h} where we sketch
the outline for two dimensional systems.

To quantify the defects we propose numerical methods that are applicable over a wide range of 
quantum phases characterized by various local and non-local orders. To this extent, we 
introduce the notion of \textit{errors} with respect to a $\textit{reference state}$ that have
been used as a  numerical probe to characterize topological orders~\cite{Jamadagni2022a, 
Jamadagni2022b}. In addition, these concepts have been employed in conjugation with machine 
learning techniques that further enhance the detection of various quantum 
phases ~\cite{Jamadagni2023}. In the current scenario, we note that the aforementioned errors 
can be interpreted as defects as in the context of QKZM. For the purpose of demonstrating the 
notion of errors with respect to a reference state, we assume the Hamiltonian, $H_{F}$ in 
Eq.~\ref{eq1} is gapped and refer to the groundstate of the Hamiltonian as the reference state. 
In a more general context, the reference state is given by the eigenstate of the operator 
corresponding to the order parameter that maximizes the same. The errors are defined by the 
action of local operators (local perturbation) on the reference state. In the following 
sections, we will explicitly introduce the errors associated with the choice of the corresponding Hamiltonian. We also note that in the rest of our description, we interchangeably use the terms defect (density) and error (density). 

\subsection{Methods to compute defect density\label{sec:def_den}}
In the following, we introduce two different computational strategies that estimate 
the density of errors. Given a gapped Hamiltonian, we compute the above based on 
expectation values of certain projectors in an exact and an approximate fashion~\cite{Han2018, Jamadagni2022b, Jamadagni2023}.

\subsubsection*{Method 1: Expectation value of the projectors}
Let us assume that the Hamiltonian, $H_F$, can be expressed as a sum of $k$-local Hamiltonians, 
$h_{i}$, as in Eq.~\ref{eq:gapped_ham},
\begin{equation}
H_0 = \sum\limits_{i}h_{i}.
\label{eq:gapped_ham}
\end{equation}
We denote the energy spectrum of the $k$-local Hamiltonian $h_{i}$ by $\lambda^{i}_{j}$ and the 
corresponding eigenstates by $\ket{\lambda^{i}_{j}}$ with the groundstate denoted by setting $j=0$. For any 
given state $\ket{\psi}$, the total number of local errors with respect to the reference state 
$\ket{\lambda^{i}_{0}}$ for the $k$-sites is given by $\eta_{i}$, as in the Eq.~\ref{eq:doe_exp},

\begin{equation}
\eta_{i} = 1-\braket{\mathcal{P}^{i}_{0}},
\label{eq:doe_exp}
\end{equation}
where $\mathcal{P}^{i}_{0} = \ket{\lambda_{0}^{i}}\bra{\lambda_{0}^{i}} $
(or in the case of degeneracy $\mathcal{P}^{i}_{0} = \sum\limits_{d}\ket{\lambda_{0d}^{i}}\bra{\lambda_{0d}^{i}}$ with $d$-degenerate 
groundstates). For $k$ sites, since
$\sum\ket{\lambda^{i}_{0}}\bra{\lambda^{i}_{0}}=\mathds{1}$,
and $\mathcal{P_{0}^{i}}$ projecting into the groundstate of the $k$-local
Hamiltonian, results in the total number of local errors, $\eta_{i}$, as in 
Eq.~\ref{eq:doe_exp}. Finally, we arrive at the total defect density as the 
total number of defects averaged over the system size, i.e.,
$\eta=\sum_{i}\eta_{i}/N$.

\subsubsection*{Method 2: Monte-Carlo based sampling}

The defect densities can be obtained by measuring the wavefunction in the error
basis. Numerically, we simulate the above by employing the Monte-Carlo sampling.
First, we compute the expectation values of the $k$-local projectors
$\mathcal{P}_{0}^{i}$. 
Next, we generate a random number, $r$ and if $r < \braket{\mathcal{P}_{0}^{i}}$
we identify the $k$-sites with a no-error configuration and otherwise 
as an error. In the case of the erroneous configuration, we apply the projector
$P_{0}^{i}$, else we apply $\mathds{1}-P_{0}^{i}$ and renormalize the wavefunction.
We repeat the above strategy over all the remaining $k$-local Hamiltonians and 
capture the corresponding errors for a single trajectory. 
We obtain the defect densities by normalizing over the system size and averaging
it over considerable number of trajectories.

We note that the above process is akin to simulating the measurement of a 
wavefunction in an experimental setup. However, in a real experiment it might 
not always be possible to engineer the $k$-local projector and  further perform a 
mutli-site measurement. In order to circumvent the above complexity, in the later 
sections, we propose a model dependent $m$-local measurement operator with $m<k$ that 
captures (partial) information about errors at the same time being experimentally 
more accessible.

\begin{figure*}[t]
\begin{center}
  \begin{tabular}{cp{0.01mm}c}
    \subfig{(a)}{\includegraphics[width=.4\linewidth]{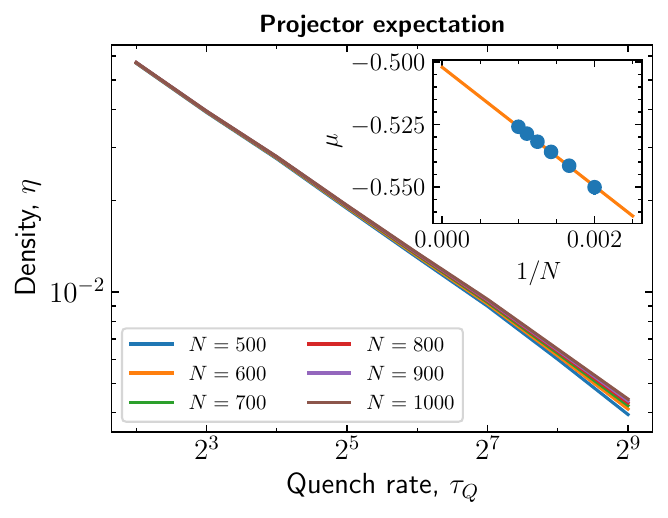}}
    &&
    \subfig{(b)}{\includegraphics[width=.4\linewidth]{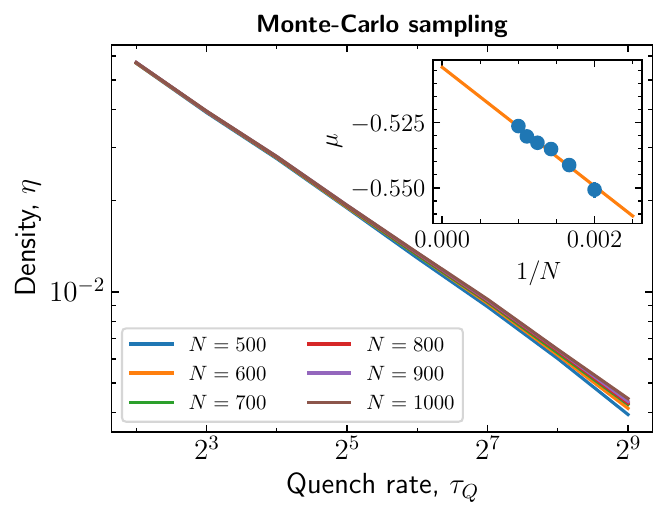}}
  \end{tabular}
\end{center}
\caption{QKZM with the quench protocol driven across a paramagnetic-ferromagnetic QPT as in the tranverse 
field Ising model. Defect density, $\eta$, as a function of the quench rate, $\tau_{Q}$, with $\eta$ 
computed using the strategy based on (a) expectation value of the projector, (b) Monte-Carlo sampling. 
The linear scaling remains indistinguishable for progressively increasing values of $N$ indicating 
convergence to the behavior in the thermodynamic limit. We note that in the rest of the figures, we 
consider the y-axis on a decimal logarithm scale for better representation but the actual fit is made 
with respect to $\log_{2}$ on both axes. (Inset) By performing finite-size scaling analysis we 
estimate the critical exponent $\mu$ to be (a) 0.502(1), and (b) 0.504(2) in the thermodynamic limit.}
\label{fig:tfim}
\end{figure*}

\subsection{Extraction of the scaling exponent\label{sec:ext_ce}}

Having defined two different strategies to compute the defect density, in the following, 
we outline a numerical procedure to determine scaling exponent, $\mu$ arising in the QKZM. 
\begin{enumerate}

 \item We set the initial time of the quench dynamics to be $t_{i} = \beta\tau_{Q}$, where 
 $\beta$ is some constant such that $g(t_{i}) \gg g(t_c)$ with the final time of the dynamics 
 as $t_{f}=0$, leading to $H_{I}$ being turned off. Next, we evolve the initial state through a 
 QPT by employing the Hamiltonian in Eq.~\ref{eq1};
 
 \item For different quench rates, we evolve the corresponding initial states. Following the 
 above, we compute the error densities of the final evolved state;   
 
 \item For a given system size, we further extract $\mu$ by linearly fitting the defect 
 density, $\eta$ with respect to the quench rate, $\tau_{Q}$, (on a $\log-\log$ scale i.e., $\log_{2}\eta 
 \propto -\mu\log_{2}\tau_{Q}$). We then perform a  finite-size scaling analysis to estimate the 
 scaling exponent $\mu$ in the thermodynamic limit.
 
\end{enumerate}

\section{Quantum phases with local order \label{sec:local_order}}

To demonstrate the protocol, in this section, we consider a setting wherein a quantum phase characterized 
by local order is driven across criticality by a time-dependent perturbation.
We begin by studying the paradigmatic model, the Transverse Field Ising Model (TFIM) that exhibits
local order. The model consists of linear chain of $N$ spin-1/2's with open boundary condition defined by 
the following Hamiltonian
\begin{equation}\label{eq:h_tfim}
 H_{\text{TFIM}}(t) = -\sum\limits_{i=1}^{N-1}\sigma_{z}^{i}\sigma_{z}^{i+1} - g(t)\sum\limits_{j=1}^{N}\sigma_{x}^{j},
\end{equation}
where the nearest neighbor interaction is ferromagnetic in nature with  the strength of transverse field being time-dependent, $g(t)$, as defined previously in Eq.~\ref{eq1}.

The above Hamiltonian exhibits a QPT with the groundstate being a paramagnet in the low perturbed regime
while being a ferromagnet in the high perturbed regimes with a criticality at $g(-\tau_{Q})$=1. 
The paramagnet-ferromagnet transition belongs to the Ising universality class with the critical exponents 
$\mu=z=1.0$ \cite{Sachdev_2011}. To estimate the critical exponent, we first choose the initial state to be the groundstate of
Eq.~\ref{eq:h_tfim} at some high field strength, thereby belonging to the paramagnetic phase. Then we evolve the system across the QPT guided by the quench protocol. In the current context, 
the reference state that leads to the construction of errors is the ferromagnetic groundstate i.e., 
$\ket{0000...00}$ and $\ket{1111...11}$. The presence of the transverse field gives rise to errors that are 
recognized by neighboring spins with opposite parity when measured in the $\sigma_z$ basis. Having 
introduced the quench dynamics and the errors associated with the ferromagnetic groundstate, we 
quantify the scaling exponent by employing the strategies as outlined in Sec.~\ref{sec:def_den}.

\subsubsection{Quantifying criticality using expectation value}

As noted in Sec.~\ref{sec:def_den}, the local defect density is captured by Eq.~\ref{eq:doe_exp}, 
which in the current scenario reduces to, $\mathcal{P}_{0}^{i} = \ket{00}\bra{00}_{(i, i+1)} + 
\ket{11}\bra{11}_{(i, i+1)}$. Equivalently, this can be represented as the expectation of the projector
$\frac{\mathds{1} - \sigma_{z}^{i}\sigma_{z}^{i+1}}{2}$, that detects the presence of domain walls.
The total error density is therefore given by
\begin{equation}\label{eq:exp_tfim}
\eta = \frac{1}{N}\sum\limits_{i=1}^{N-1}\left(1 - \braket{\ket{00}\bra{00}}_{(i, i+1)} - \braket{\ket{11}\bra{11}}_{(i, i+1)}\right).
\end{equation}

To extract the exponent, $\mu$, for a given system size $N$, we deploy the procedure as outlined in 
Sec.~\ref{sec:ext_ce}. In order to gain access to significantly higher system sizes we consider the Matrix Product 
State (MPS) representation of the quantum states in the rest of the analysis. The initial state
i.e., the ground state is computed using the Density Matrix Renormalization Group (DMRG) algorithm
and the time evolution is performed using the Time Evolution Block Decimation (TEBD) 
algorithm~\cite{Vidal2004} by choosing the Trotter scheme wherein the total error scales 
quadratically in the time step. We note that both the above implementations are realized by 
employing the ITensor library~\cite{itensor2022}. From the numerical simulations, see 
Fig.~\ref{fig:tfim}, we obtain the critical exponent to be $\mu=0.502(1)$ that agrees well with 
the results obtained in Refs.~\onlinecite{sachdev1999quantum, Keesling_2019}.
We note that the 
error can be further suppressed by choosing higher order Trotterization schemes.

\subsubsection{Quantifying criticality using Monte-Carlo sampling}
In the following, we employ the Monte-Carlo based sampling to obtain the errors using single-site
measurements as in Ref.~\onlinecite{Jamadagni2023}. That is, we sample the final evolved wavefunction in 
the $\sigma_{z}$ basis and identify the errors by the presence of different parity bits on the neighboring 
sites. In other words, the simulation of Monte-Carlo sampling generates the so called shot data as in the 
context of quantum computing. The tensor network based simulation in conjugation with the Monte-Carlo
sampling leads to the critical exponent to be $\mu=0.504(2)$ which is in good agreement with the value
obtained earlier. The projector, $\mathcal{P}_{0}^{i}$, in the TFIM is a diagonal operator and 
therefore is easier to access in an experimental setting. However, in the next sections, we explore
systems that involve non-diagonal projection operators. Thereby, Monte-Carlo sampling techniques introduced 
here provide a means to estimate the defect density experimentally in a feasible manner.

\section{Quantum phases with symmetry protected topological order \label{sec:spt_order}}
In this section, we extend the analysis to the context of topological phases, phases that are
beyond the Landau symmetry breaking principle thereby being characterized by non-local order parameters.
We restrict our analysis to topological phases hosting short-range entangled states with a given symmetry, 
also know as Symmetry Protected Topological (SPT) states. The short-range entanglement implies that
they can smoothly deformed into a product state unless the deformation preserves the symmetry.
In other words, SPT states cannot be mapped to a product state using finite-depth symmetry preserving 
local unitaries~\cite{Chen2010}. 
In the following, we consider three different time-dependent Hamiltonians that host SPT phases along with a quench 
protocol that drives these phases across a QPT. We further employ the computational methods introduced earlier 
to estimate the scaling exponent associated with the topological phase transition. 

\begin{figure}[t]
\begin{center}
\includegraphics[width=0.9\linewidth]{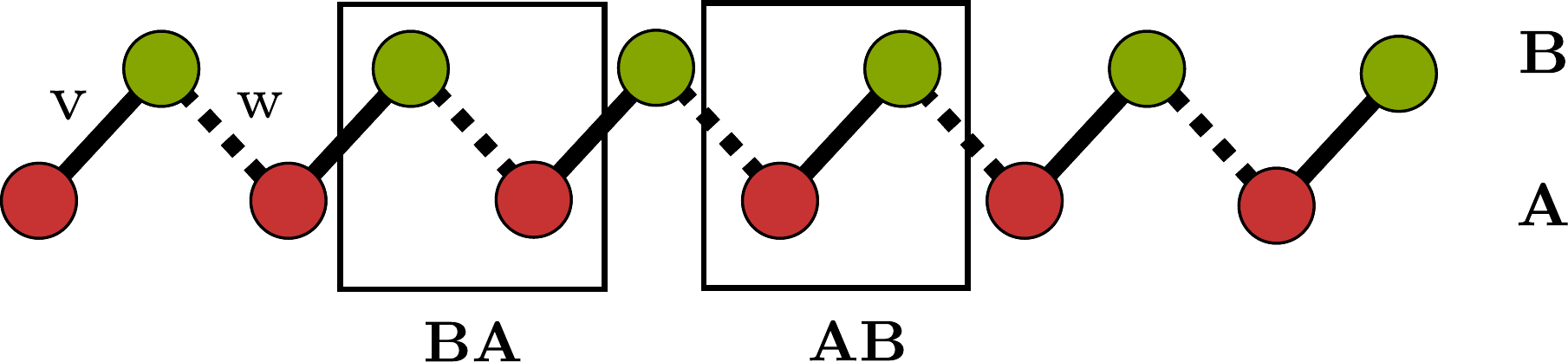}
\vspace{0.75cm}
\caption{The SSH model describes the hopping of a particle between distinct neighboring sites A and B 
with alternating bond strengths $v$ and $w$ identified by the unit cells AB and BA. In the hardcore boson 
variant, we consider a half-filled lattice of $N$ sites with each unit cell hosting a single particle in 
the limits of $v=0$ and $w=0$. We note that when $v=0$ we observe $N/2$ dimerized unit cells while in the 
other limit $w=0$ we notice $N/2-1$ dimers in addition to unpaired edge sites giving rise to the so called 
edge modes.}
\label{fig:ssh_lattice}
\end{center}
\end{figure}

\begin{figure*}[t]
\begin{center}
  \begin{tabular}{cp{0.01mm}c}
    \subfig{(a)}{\includegraphics[width=.4\linewidth]{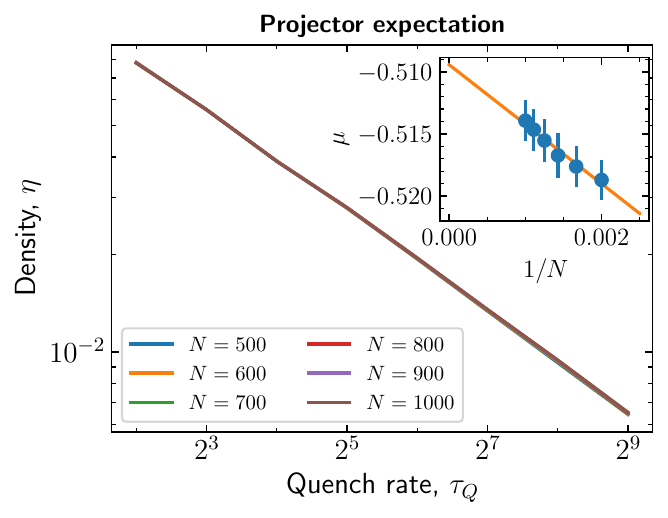}}
    &&
    \subfig{(b)}{\includegraphics[width=.4\linewidth]{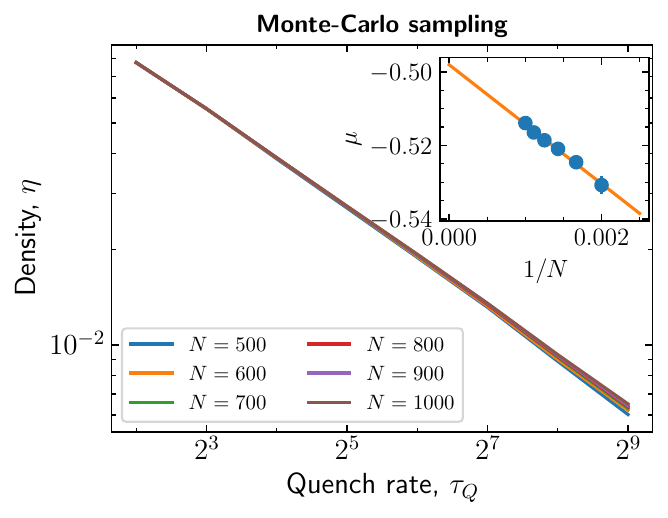}}
  \end{tabular}
\end{center}
\caption{Quench dynamics crossing a QPT involving symmetry protected topological phases as in the SSH model.
Numerically obtained defect density scales linearly with the quench rate with the former computed using
(a) expectation value of the projector, (b) Monte-Carlo sampling. (Inset) Finite-size scaling analysis
to obtain the scaling exponent, $\mu$=(a) 0.509(1) and (b) 0.498(2).}
\label{fig:ssh_mu}
\end{figure*}

\subsection{Su–Schrieffer–Heeger (SSH) model}
The SSH model was introduced in the context of a particle hopping on 1D-lattice~\cite{Su1980} and
is known to host phases that exhibit SPT order. We consider a time-dependent hardcore bosonic version 
of the above model, whose Hamiltonian is described by the following,
\begin{align}\label{eq:td_ssh_ham}
H_{\text{SSH}}(t) = v(t)\sum\limits_{i=1}^{N/2}\sigma_{-}^{2i-1}\sigma_{+}^{2i} 
+ w\sum\limits_{i=1}^{N/2-1}\sigma_{-}^{2i}\sigma_{+}^{2i+1} + \text{h.c.,}
\end{align}
with $v(t)/w = -t/\tau_Q$, $t\in(-\infty, 0]$ with $\tau_Q$ being the quench rate. The time independent 
version of Eq.~\ref{eq:td_ssh_ham} with $v(t)=v$ is exactly solvable in the case of periodic boundary 
conditions. It hosts gapped phases in the extremal limits of $v \ll w$ and $v \gg w$ with the gap closing 
at $v=w$. In the case of open boundaries, in the limit of $v\ll w$ it is known that the topological phase 
is identified by the presence of edge modes, characterized as non-trivial SPT phase. While in the other 
limit the phase remains topological with no edge modes, characterized as trivial SPT phase with the phase 
transition occurring at $v=w$~\cite{Elben2020,Jamadagni2022b}. For the rest of the analysis we consider 
the SSH Hamiltonian on a 1D-chain i.e., with open boundaries. 

The quench protocol to validate QKZM involves driving an initial state belonging to the trivial SPT 
phase i.e., the groundstate of Hamiltonian in Eq.~\ref{eq:td_ssh_ham} at some $v \gg w$. With the 
final state belonging to the non-trivial SPT phase, we set the reference state as groundstate of the 
above Hamiltonian at $v=0$, given by the singlet configuration in each of the BA unit cells i.e.,

\begin{equation}\label{eq:ssh_ham_gs}
 \ket{\phi}_{\text{BA}} = \frac{1}{\sqrt{2}}\prod_{i\in B}\left(\ket{0}_{i}\ket{1}_{i+1} - \ket{1}_{i}\ket{0}_{i+1}\right),
\end{equation}
where BA unit cells are as in Fig.~\ref{fig:ssh_lattice}. Therefore, deviations from the 
singlet configuration in Eq.~\ref{eq:ssh_ham_gs}
give rise to local errors in the corresponding unit cell, that are characterized as density fluctuations, 
$\ket{\bf{0}} = \ket{00}$, $\ket{\bf{1}} = \ket{11}$ and phase fluctuations,  
$\ket{\pmb{\plus}} = \frac{1}{\sqrt{2}}(\ket{01} + \ket{10})$.

In the following, we compute the error density and determine the critical exponent by employing 
techniques as outlined in earlier sections.  We assume the state mentioned in Eq.~\ref{eq:ssh_ham_gs} as the reference state and first estimate the local error density by setting $\mathcal{P}_{0}^{i} =  \ket{\pmb{\minus}}\bra{\pmb{\minus}}$ 
in Eq.~\ref{eq:doe_exp}. This further results in the total defect density given by
\begin{equation}
\eta = \frac{1}{N}\sum_{i\in B}\left(1-\braket{\ket{\pmb{\minus}}\bra{\pmb{\minus}}}_{(i,i+1)}\right),
\label{eq:den_sshni}
\end{equation}
which is due to the fact that each unit cell satisfies the following relation:
$\sum\limits_{\lambda}\ket{\lambda}\bra{\lambda} = \mathds{I}$ for $\lambda \in \{\bf{0}, \bf{1}, 
\pmb{\plus}, \pmb{\minus}\}$. Further, by computing the defect density using the above equation, 
we estimate the scaling exponent as $\mu=0.509(1)$, see Fig.~\ref{fig:ssh_mu}.

Furthermore, it is also possible to determine the critical exponent by employing Monte-Carlo sampling.
To this extent, we sample the final evolved state in the excitation basis 
given by \{$\ket{\bf{0}}$, $\ket{\bf{1}}$, $\ket{\pmb{\plus}}$, $\ket{\pmb{\minus}}$\} with respect to the 
earlier defined reference state, see Eq.~\ref{eq:ssh_ham_gs}. 
With the total number of errors given by the sum of density fluctuations, $\{\bf{0}, \bf{1}\}$ and phase fluctuations, 
$\{\pmb{\plus}\}$ results in the critical exponent $\mu=0.498(2)$ shown in Fig.~\ref{fig:ssh_mu}. 
The trival-SPT transition of the SSH model belongs to BDI universality class with the critical exponents 
$\mu=z=1.0$ \cite{Ryu_2010, Velasco_2017, Chen_2017}. Therefore, our method is capable of estimating the expected critical exponent (1/2 for this case) across a topological phase transition~\cite{Sun2022}.

\subsection{Extended SSH model}
\begin{figure*}
\begin{center}
\includegraphics[width=0.95\linewidth]{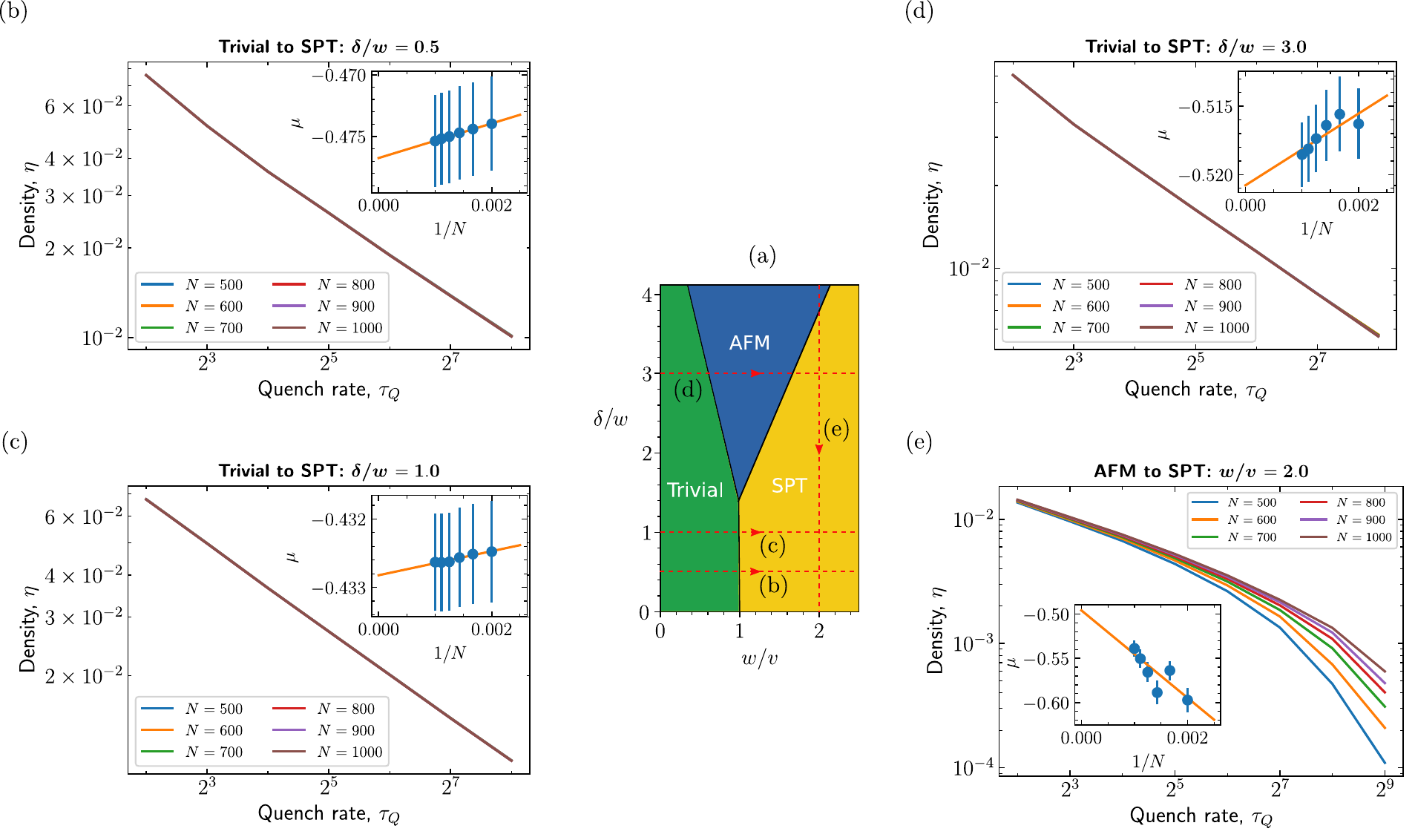}
\caption{(a) Qualitative groundstate phase diagram of the extended SSH model with arrows indicating 
the quench dynamics through different QPTs (b, c, d) from trivial to SPT phase by fixing $\delta/w$ to be
(b) 0.5, (c) 1.0, (d) 3.0, and (e) from antiferromagnetic (AFM) to SPT phase by fixing $w/v=2$.
In the former, the linear scaling of defects (on a $\log-\log$ scale) is in accordance with the QKZM (inset) 
resulting in the critical exponent, $\mu$ to be (b) 0.4767(1), (c) 0.4328(1), and (d)0.521(1). (e) The finite
size effects in the AFM-SPT quench are more predominant and are reflected in the deviations from the linear scaling included in the error-estimation of (inset) critical exponent, $\mu$=0.50(2).}
\label{fig:essh}
\end{center}
\end{figure*}

In this section, we extend the analysis to the case of the extended SSH model whose Hamiltonian is given by  
\begin{align}
 \begin{split}
  H_{\text{eSSH}} &= \frac{v}{2}\sum\limits_{i=1}^{N/2}\sigma_{x}^{2i-1}\sigma_{x}^{2i} + \sigma_{y}^{2i-1}\sigma_{y}^{2i} + \delta\sigma_{z}^{2i-1}\sigma_{z}^{2i} \\
         &+ \frac{w}{2}\sum\limits_{i=1}^{N/2-1}\sigma_{x}^{2i}\sigma_{x}^{2i+1} + \sigma_{y}^{2i}\sigma_{y}^{2i+1} + \delta\sigma_{z}^{2i}\sigma_{z}^{2i+1}. \\
 \end{split}
 \label{eq:h_issh}
\end{align}
We note that setting $\delta=0$ recovers the SSH model discussed in the previous section. The phase diagram 
of the extended SSH model has been discussed in Refs.~\onlinecite{Elben2020, Jamadagni2022b, Jamadagni2023} 
and is known to host trivial and non-trivial SPT phases along with an antiferromagnetic (AFM) phase (as 
$\delta \rightarrow \infty$), see Fig.~\ref{fig:essh}, for a qualitative sketch of the same. Given the 
rich phase diagram, we obtain the associated critical exponents by driving across various QPTs. To this 
extent, we consider two time-dependent variants of the Hamiltonian in Eq.~\ref{eq:h_issh} where: 
(i) $v$ is replaced by a time-dependent function, with $\delta$ and $w$ being a constant, 
(ii) $v$ and $w$ remain a constant, with $\delta$ being replaced by a time-dependent function. 
We proceed by presenting a quantitative analysis of the critical behavior using the expectation value of the 
groundstate projector, while the Monte-Carlo sampling approach has been analyzed in 
App.~\ref{app:A}.

\subsubsection{Trivial to SPT transition\label{issh_qp1}}
We dynamically traverse across the trivial to non-trivial SPT topological transition by employing a quench 
protocol i.e., by setting $v$ to be $v(t) = -t/\tau_{Q}$, $ t \in (-\infty, 0]$ and $\delta$ to be a constant
belonging to the set $\{0.5, 1, 3\}$ in Eq.~\ref{eq:h_issh}. The initial state is chosen as the groundstate of 
the Hamiltonian in the regime of $v \gg w$ with the final evolved state belonging to the non-trivial SPT phase.
The reference state is as introduced in the earlier section via Eq.~\ref{eq:ssh_ham_gs}.
The above setting leads us to a final evolved state with total defect density given by $\eta$, as in Eq.~\ref{eq:den_sshni}. In Fig.~\ref{fig:essh}, we estimate the critical 
exponents corresponding to $\delta$ in $\{0.5, 1, 3\}$. Our numerical experiments show that 
with interactions turned on, the value of the scaling exponent deviates significantly
from 0.5 as in the non-interacting case. We also notice that the value decrease along the
transition line $w/v$=1 upto the point where the three phases co-exist and then recovers 
to 0.5 beyond that. To the best of our knowledge, our studies is one of the first to report such 
estimates.

\subsubsection{AFM to SPT transition}
The phase diagram allows for the exploration of critical exponent associated with the 
AFM-SPT phase transition. To this end, the quench protocol is defined by
$\delta(t)=-t/\tau_{Q}$ with $t \in (-\infty, 0]$ while fixing $w/v=2$ in Eq.~\ref{eq:h_issh}. 
The initial state for evolution is chosen to be the groundstate of $H_{\text{eSSH}}$ in the limits of 
$\delta \gg w/v$, thereby being smoothly connected to an antiferromagnet. As the system is driven
into a non-trivial SPT phase, the reference state remains the same. It might
be intuitive to conclude that the total error density is given by Eq.~\ref{eq:den_sshni}.
However, on the contrary this is not the case as there are additional corrections involved.
It is important to note that the final evolved state encapsulates the errors
of the groundstate at finite $w/v$ that need to be subtracted from Eq.~\ref{eq:den_sshni}
to obtain the accurate defect densities. To be more precise, let us denote the 
final evolved state by $\ket{\psi}_{f}$ and the ground state at $w/v$ = 2 by $\ket{\psi}_{g}$, 
with the corresponding defect density given by $\eta_{f}$ and $\eta_{g}$ respectively. 
The true defect density involving the additional corrections terms is therefore given 
by the following
\begin{equation*}
\eta_{\text{eff}} = \frac{1}{N}\sum_{i\in B}\big|\braket{\psi_{g}|\pmb{\minus}}_{(i,i+1)}\big|^2-\big|\braket{\psi_{f}|\pmb{\minus}}_{(i,i+1)}\big|^2.
\end{equation*}
It is clear to see that $\eta_{\text{eff}}$ is non-negative as the overlap of the $\ket{\psi_{g}}$ 
with the reference state $\ket{\pmb{\minus}}$ is more in comparison to the final evolved state, 
$\ket{\psi_{f}}$ that involves additional excitations. In Fig.~\ref{fig:essh}, we  note that by 
incorporating the additional correction terms we substantiate the predictions as in the QKZM and 
further obtain the scaling exponent.

\subsection{Cluster state model}
One other paradigmatic model that hosts SPT phase is the 1D cluster state model, whose Hamiltonian in the 
presence of time-dependent perturbation is given by

\begin{align}
 H_{\text{CS}} = -\sum\limits_{i=1}^{N-2}\sigma_{z}^{i}\sigma_{x}^{i+1}\sigma_{z}^{i+2} - g(t)\sum\limits_{j=1}^{N}\sigma_{x}^{j},
 \label{eq:cs_ham}
\end{align}
where $g(t)$ is as defined in Eq.~\ref{eq1}. For the rest of the analysis, we consider the Hamiltonian on a 1D 
spin chain i.e., with open boundary conditions. In the case of time-independent perturbation, the 
Hamiltonian hosts a SPT phase at low perturbation strength and a paramagnetic phase at high perturbation 
strength with a QPT  at some perturbation strength~\cite{Cong2019}. In the absence of perturbation, the
groundstate of the Hamiltonian is short-range entangled and protected by a $\mathbb{Z}_{2} \times 
\mathbb{Z}_{2}$ symmetry~\cite{Son2011}. The groundstate is also referred to as 1D cluster state 
with applications in measurement-based quantum computing~\cite{Nielsen2006,Deng2017}.

\begin{figure}[t]
 \begin{center}
\includegraphics[width=0.9\linewidth]{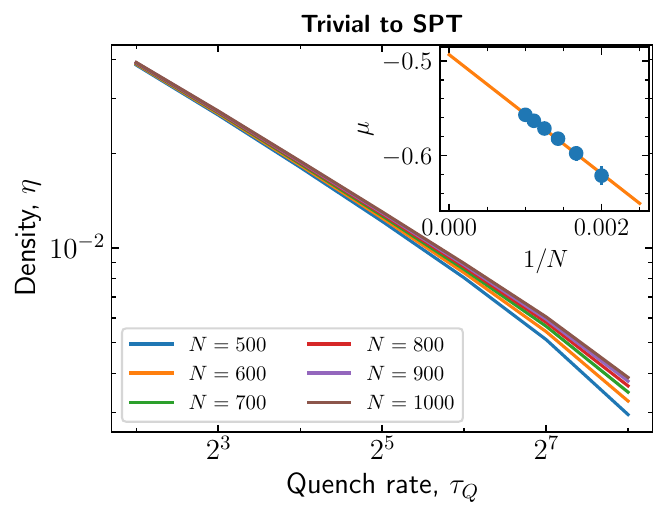}
\vspace{0.75cm}
  \caption{Steering across a trivial-SPT phase transition with the latter characterized by the cluster state 
  Hamiltonian. The predictions of the QKZM remain valid as the logarithm of the defect densities scale linearly with 
  the quench rate in $\log$ scale. We attribute minor deviations from linearity to the finite-size of the system 
  resulting in the (inset) critical exponent, $\mu$=0.493(2).} 
\label{fig:zxz_expect}
\end{center}
\end{figure}

The quench dynamics is performed by choosing the initial state to be the groundstate of the 
Hamiltonian belonging to the trivial phase. Further, we evolve the above state to a final time wherein the 
perturbation is completely turned off resulting in a final state belonging to the SPT phase. We further 
compute the local error density of the final evolved state by setting 
$\mathcal{P}_{0}^{i} = \frac{\mathds{1} + \sigma_{z}^{i}\sigma_{x}^{i+1}\sigma_{z}^{i+2}}{2}$ in 
Eq.~\ref{eq:doe_exp}. $\mathcal{P}_{0}^{i}$ projects a given quantum state into the groundstate of the 
cluster state Hamiltonian, thereby resulting in the total defect density given by
\begin{equation*}
    \eta = \frac{1}{N}\sum\limits_{i=1}^{N-2}\left(1 - \braket{\frac{\mathds{1} + \sigma_{z}^{i}\sigma_{x}^{i+1}\sigma_{z}^{i+2}}{2}}\right).
\label{eq:den_cs}
\end{equation*}
Our numerical analysis shows the critical exponent to be $\mu=0.493(2)$, see Fig.~\ref{fig:zxz_expect}. 
This establishes that the numerical methods introduced here work for wider class of gapped phases that demand multi-site interactions. Estimates for the critical exponents are obtained for different perturbed model involving the cluster state Hamiltonian leading to the same scaling exponent as in our case \cite{Son_2011}.

\section{Quantifying criticality using digital quantum computers\label{sec:cric_qc}}

\begin{figure*}
\begin{center}
\includegraphics[width=0.875\linewidth]{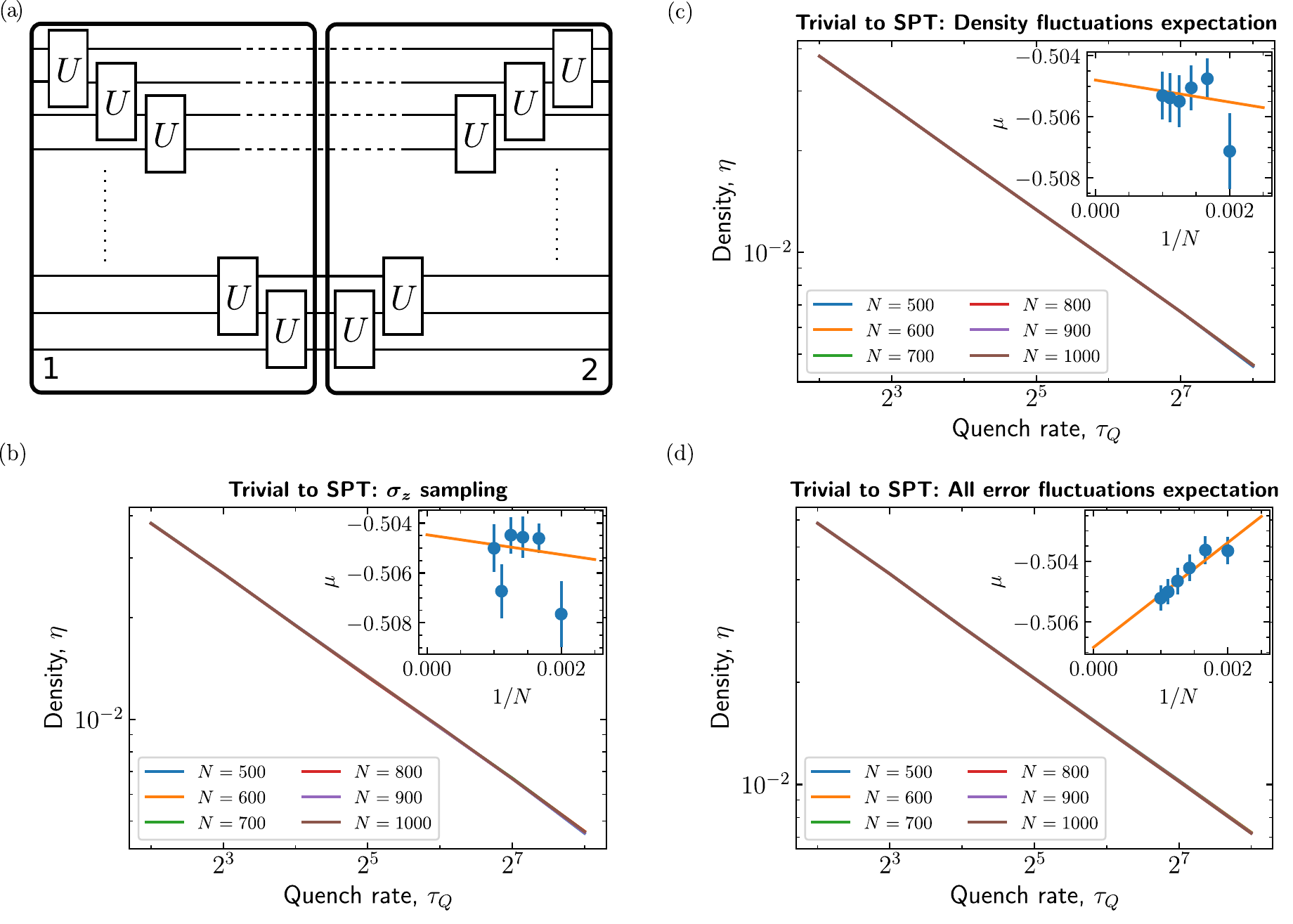}
\caption{Simulating the quench dynamics involving a topological phase transition as in the extended SSH 
model on a digital quantum computer. (a) The TEBD protocol comprising of the two-qubit unitary, $U$, see 
Eq.~\ref{eq:trotter_exp}. The decomposition of $U(\alpha, \beta, \gamma)$ in terms of single qubit 
rotations and CNOTs is as in Refs.~\onlinecite{Vatan2004,Smith2019}.
The predicted power law scaling in the QKZM is recovered by considering (b) partial defect density i.e., 
density fluctuations estimated by sampling in $\sigma_z$ basis similar to shot data in experiments, (c) 
partial defect density by computing the expectation value of the density fluctuations, and (d) total 
defect density determined using the projector expectation value. (Inset) The scaling exponent, $\mu$ 
obtained by performing finite size scaling resulting in (b) 0.504(2), (c) 0.505(1), and (d) 0.507(1) that 
are in good agreement among themselves while also to the quench protocol employed in Sec.~\ref{sec:spt_order}}. 
\label{fig:issh_qc}
\end{center}
\end{figure*}

The recent advancement in quantum hardware has enabled the exploration of several quantum many-body 
phenomena~\cite{Smith2019,deLeseleuc2019,Semeghini2021,Satzinger2021,iqbal2023,Li2023} 
using Noisy Intermediate Scale Quantum (NISQ) devices~\cite{Preskill2018}. QKZM has been validated on 
both analog~\cite{Keesling_2019} and digital architectures~\cite{Dupont2022}. The former provides
efficient implementation of quench dynamics while the latter is applicable in a more general setting.
For instance, a digital based experiment as well as intensive numerical investigation of the QKZM in TFIM was 
explored in recent work, see Ref.~\onlinecite{Dupont2022}. Motivated by the above work, we slightly alter our 
quench protocol that maps two Hamiltonians, $H_{I}$ and $H_{F}$, as follows 
\begin{equation}
H(t) = (1-t/\tau_{Q})H_{I} + (1+t/\tau_{Q})H_{F}.
\label{eq:h_tdhahb}
\end{equation}
We emphasize our method is capable of estimating the critical exponent in models with quantum phases that are beyond
the Landau symmetry breaking principle. To exemplify, in this section, we supplement the analysis to the case of the extended SSH Hamiltonian. To this extent, we map the hopping terms in Eq.~\ref{eq:h_issh} with strengths $v$ and $w$ to $H_{I}$ and $H_{F}$ respectively resulting 
in
\begin{align}
 \begin{split}
  H_{\text{eSSH}}(t, \delta) &= \frac{v'(t)}{2}\sum\limits_{i=1}^{N/2}\sigma_{x}^{2i-1}\sigma_{x}^{2i} + \sigma_{y}^{2i-1}\sigma_{y}^{2i} + \delta\sigma_{z}^{2i-1}\sigma_{z}^{2i} \\
         &+ \frac{w'(t)}{2}\sum\limits_{i=1}^{N/2-1}\sigma_{x}^{2i}\sigma_{x}^{2i+1} + \sigma_{y}^{2i}\sigma_{y}^{2i+1} + \delta\sigma_{z}^{2i}\sigma_{z}^{2i+1}, \\
 \end{split}
 \label{eq:h_sshi}
\end{align}
with $v'(t)=1-t/\tau_{Q}$, $w'(t)=1+t/\tau_{Q}$, $t \in [-\tau_{Q}, \tau_{Q}]$ and set $\delta=3$ to be
a constant, see Fig.~\ref{fig:essh}(a). The initial state that is evolved is identified by the singlet 
configuration in the AB unit cells  i.e., a state belonging to the trivial SPT phase (groundstate of 
$H_{\text{eSSH}}(-\tau,3)$). The key advantage of using such a protocol is that initial state 
can be expressed analytically. That is, the groundstate can be expressed as a tensor product of the
2-qubit Bell state ($\frac{1}{\sqrt{2}}(\ket{01}-\ket{10})$) that can be further prepared by using a sequence 
of single and two qubit gates. With the dynamics driving the above state into a non-trivial 
SPT phase, the reference state is set to the singlet configuration in the BA unit cells. To simulate the 
dynamics, as earlier we employ the TEBD protocol, see Fig.~\ref{fig:issh_qc}(a) that involves a two-qubit 
unitary, $U$, of the form
\begin{equation}
U = \text{exp}[i(\alpha\sigma_{x}\otimes\sigma_{x} + \beta\sigma_{y}\otimes\sigma_{y} + \gamma\sigma_{z}\otimes\sigma_{z})].
\label{eq:trotter_exp}
\end{equation}
The above can be realized on a digital quantum architecture by decomposing it further
into single-qubit Pauli rotations and two-qubit Controlled-NOT (CNOT) gates~\cite{Vatan2004,Smith2019}, 
We note that the procedure outlined in this section and the simulations thereof, 
assume an ideal quantum computing architecture. Relaxing the above constraints involves employing additional techniques 
for instance, fine-tuned finite-size scaling analysis, employing circuit depth reduction techniques to achieve the 
required evolution in shorter time, and integrating error mitigation strategies, to mention a few. We leave this 
exploration in the context of topological phases to the future.

In the following, by employing different methods we estimate the defect density. Further, we establish that
the scaling exponent determined using partial defect density is in good congruence with that obtained
using the total defect density. Importantly among the above methods, the former remains more accessible in an 
experimental setup.

\begin{figure*}
\begin{center}
\includegraphics[width=0.875\linewidth]{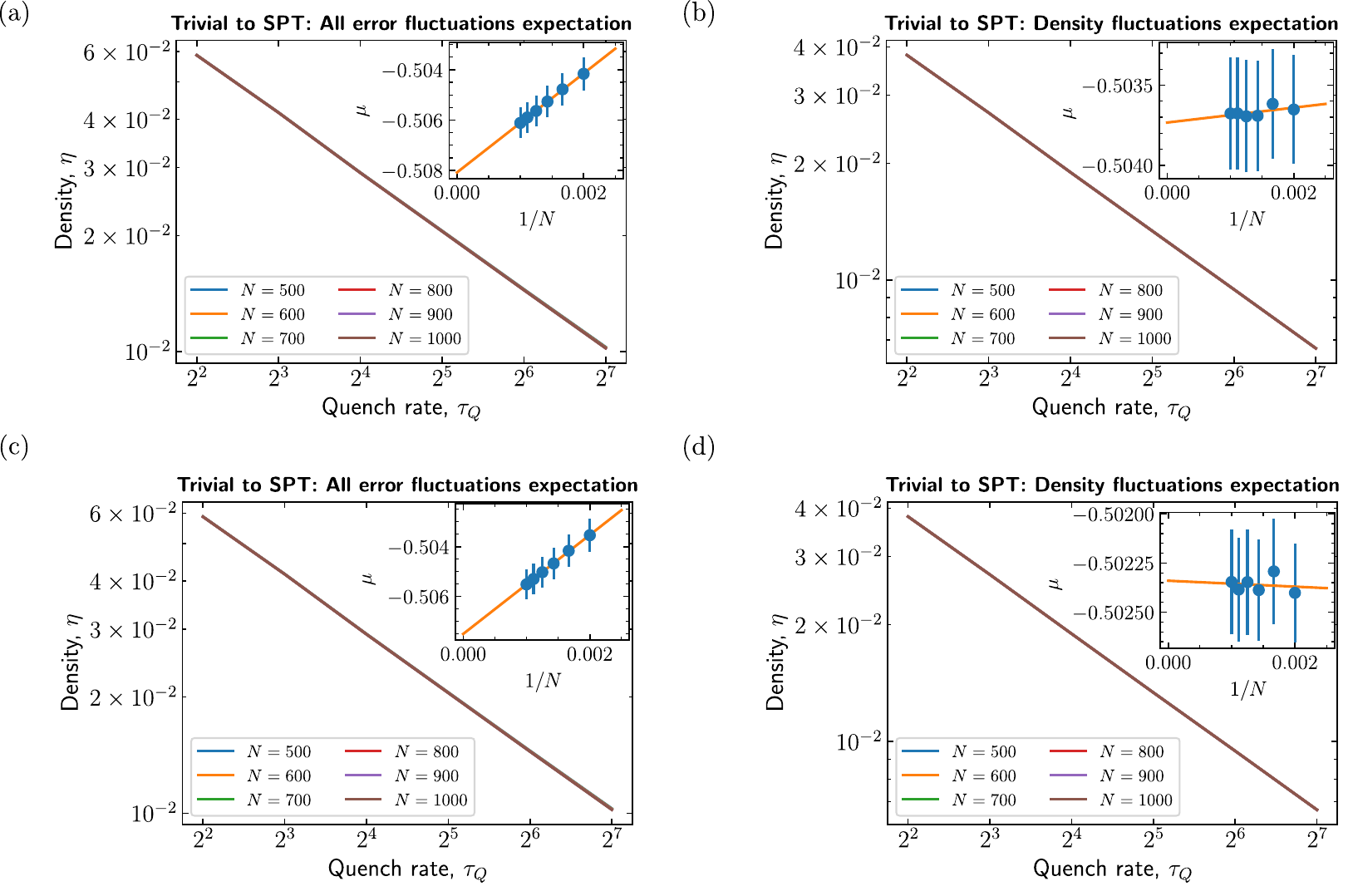}
\caption{Estimating the scaling exponent by introducing SWAP noise. We recover 
the power law scaling and the associated exponent in the case of half-swap noise by computing 
the defect density using projector expectation value of (a) all errors and (b) density fluctuations
resulting in $\mu=$ (a) 0.5081(1) (b) 0.5037(5) respectively. Similarly, in the context of full-swap 
noise, the predictions of QKZM remains valid as we recover the scaling exponent as in the noiseless 
evolution by computing the defect density using projector value of (c) all errors and (d) density 
fluctuations resulting in $\mu=$ (c) 0.5075(3), (d) 0.5023(8) respectively.}
\label{fig:issh_swap_injection}
\end{center}
\end{figure*}

\subsection{Simulating local measurements}
In the earlier sections, we have introduced strategies to compute the defect densities based
on the expectation value of certain projectors. These projectors in the case of the (extended)
SSH model involve two-sites thereby requiring two-qubit measurements that are relatively 
difficult to realize on digital computing architectures. We relax the  above requirement
by measuring the final evolved wavefunction belonging to the non-trivial SPT phase in 
$\sigma_{z}$ basis alone. As noted earlier, this refers to the shot data in a real
experiment setting and can be numerically simulated by employing the Monte-Carlo sampling
in the $\sigma_{z}$ basis. However, we note that from the above measurement data it is 
possible to partially capture the error density in terms of the density fluctuations
with no access to phase fluctuations due to the single-qubit measurement. Crucially,
the partial error density still scales linearly with the quench rate 
(on the $\log-\log$ scale) further validating the prediction of the QKZM. The critical 
exponent in the current scenario is given by $\mu=0.509(3)$ as shown in Fig.~\ref{fig:issh_qc}
which is in good agreement with the one obtained earlier in Fig.~\ref{fig:essh}(d). From 
our numerical experiments, we conclude that single qubit measurements suffice to 
estimate the critical exponent in the case of topological phase transitions as in the
extended SSH model.

\subsection{Computing partial and total defect densities}
To cross-validate the above results, we compute the expectation value of the final evolved state
with respect to different error projectors that are given by 
$\{\ket{\pmb{0}}\bra{\pmb{0}}, \ket{\pmb{1}}\bra{\pmb{1}}, \ket{\pmb{\plus}}\bra{\pmb{\plus}}\}$.
The total defect density is the sum of the expectation values of the above projectors resulting in 
Eq.~\ref{eq:den_sshni}. One other quantity that provides an approximation is the partial defect 
density given by considering only the projectors of density fluctuations similar to the simulated
local measurements.

By considering the total defect density, our numerical simulation shows the scaling exponent
to be $\mu=0.508(1)$, while scaling analysis of partial defect density results in $\mu=0.507(1)$, see
Fig.~\ref{fig:issh_qc}. We highlight the fact that the partial defect density has substantially higher 
errors as in Fig.~\ref{fig:issh_qc}(c, d) while performing the finite-size scaling analysis in 
comparison to that of the total defect density as in Fig.~\ref{fig:issh_qc}(e). However, the 
values of the scaling exponent obtained in three cases agree upto three decimals.

\subsection{Estimating critical exponent under error invariant noisy evolution}
In this section, we show that access to the errors of gapped phases allow the exploration of noisy 
dynamics (in this case coherent noise) that respect the QKZM. In the current scenario, we note that the 
gapped errors given by $\{\ket{\pmb{0}}, \ket{\pmb{1}}, \ket{\pmb{\plus}}\}$ remain invariant under the 
action of a SWAP gate. This further leads us to the notion that dynamics of the extended SSH model driven 
across a trivial non-trivial topological phase interspersed with SWAP unitaries leave the scaling exponent 
invariant as in the noiseless scenario thereby recovering the prediction of the QKZM. To validate the 
above, we consider the evolution as in the TEBD protocol and inject SWAPs unitaries leading to noisy 
evolution. We study two different SWAP injection protocols and label them as half-swap and full-swap 
injection. To motivate the above protocols we consider the TEBD algorithm as in Fig.~\ref{fig:issh_qc}(a) 
with a half (full)-swap injection implying a SWAP is prepended to all the unitaries in block identified by 
2 (1 and 2 blocks). In both the scenarios, we recover the scaling exponent as in the noiseless evolution 
obtained by computing the expectation value as in Eq.~\ref{eq:den_sshni}, see 
Fig.~\ref{fig:issh_swap_injection}(a), (c). In addition, we also note that this can be achieved on a 
digital quantum computer as the density fluctuations i.e.,  shot measurement in the $\sigma_{z}$ basis, 
also recover the expected scaling exponent, see Fig.~\ref{fig:issh_swap_injection}(c), (f). In summary, we 
observe that unitaries that leave the error space invariant, when injected into the quench dynamics retain 
the prediction of the QKZM thereby leading to same scaling exponent as in the absence of unitary injection.




\section{Summary and Discussion \label{sec:summary}}
In summary, we have briefly reviewed the QKZM that establishes a relationship between defect density and quench 
rates via a critical exponent when a quantum system is driven across a QPT. In the 
current scenario, in the context of gapped quantum phases, we recast the notion of errors with respect to a 
given reference state as defects. Further, we computed the defect density based on the expectation value of
the projection operators with respect to a predefined reference state. The values from the exact computation
provide a means to estimate the critical exponent numerically. In addition, we showed that Monte-Carlo based
sampling, akin to measurement in real experiments, provides an alternative to determine the critical exponent.

We adopted the introduced computational strategies to different spin models with QPTs involving local 
and topological orders. To this extent, we reproduced the scaling
critical exponent in the TFIM and SSH models while effectively estimating the same in the extended 
SSH model and the cluster state model. Towards the end, we proposed a strategy to determine the scaling 
critical exponent on a digital quantum computer. As an illustration, we have considered the extended SSH 
model across a QPT involving topological phases. 

It is important to note that the methods investigated in this work, can be extended to a wide range
of gapped quantum phases. Possible future applications of the computational strategies developed in the 
current context could include the exploration of (a) QPTs involving intrinsic topological order driven 
by external fields~\cite{Jamadagni2018,Jamadagni2021}, (b) landscape of QKZM in the context of open 
quantum systems and the phase transitions thereof~\cite{Jamadagni2021}, and (c) possible relations 
between QKZM and measurement induced entanglement phase transitions~\cite{Lavasani2021a,Lavasani2021b}. 
It would also be interesting to quantify criticalities associated with QPTs involving other gapped 
quantum phases that are easily realizable on upcoming quantum hardware platforms, further allowing us to 
benchmark their performance.
\vspace{0.5cm}
\begin{acknowledgments}

AJ would like to thank Andreas L\"auchli for fruitful discussions. AB would like to thank the FISPAC Research Group, Department of Physics, University of Murcia, especially, Jose J. Fernández-Melgarejo, for hospitality during the course of this work. AB is supported by the Mathematical Research Impact Centric Support Grant (MTR/2021/000490) by the Department of Science and Technology Science and Engineering Research Board (India) and the Relevant Research Project grant (202011BRE03RP06633-BRNS) by the Board Of Research In Nuclear Sciences(BRNS), Department of Atomic Energy (DAE), India. AB also acknowledge associateship program of Indian Academy of Science, Bengaluru. We have also uploaded sample code snippets at the following repository that enables the reproduction of the results: \href{https://github.com/amitjamadagni/KZ_qdyn}{https://github.com/amitjamadagni/KZ\_qdyn}.
\end{acknowledgments}

\bibliographystyle{myaps}
\bibliography{main_arxiv}

\appendix
\section{Monte-Carlo sampling of errors in the extended SSH model\label{app:A}}

We consider the time-dependent extended SSH Hamiltonian, as in Eq.~\ref{eq:h_issh} 
and set $v(t) = -t/\tau_{Q}$, $t \in (-\infty, 0]$ i.e., we drive an initial state 
belonging to the trivial topological phase to a final state in the non-trivial topological phase across the
topological phase transition. In Sec.~\ref{issh_qp1}, we determine the criticality by considering the defect
density obtained using the strategy based on expectation values of projectors, while in this section, we 
estimate the criticality by employing the Monte-Carlo method, for the case of $\delta=1$. The 
main motivation is to benchmark and compare the scaling exponent obtained using only the density 
errors and only the phase error with that of the full error profile.

\begin{figure*}[h!]
\begin{center}
  \begin{tabular}{cp{0.01mm}cp{0.01mm}c}
    \subfig{(a)}{\includegraphics[width=.3\linewidth]{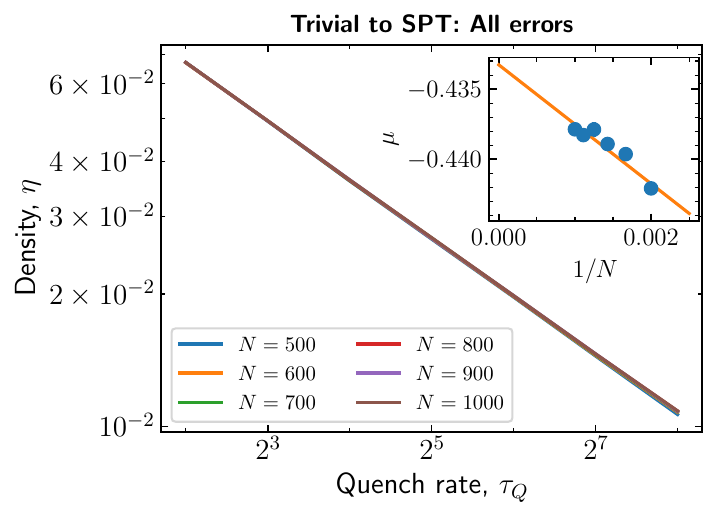}}
    &&
    \subfig{(b)}{\includegraphics[width=.3\linewidth]{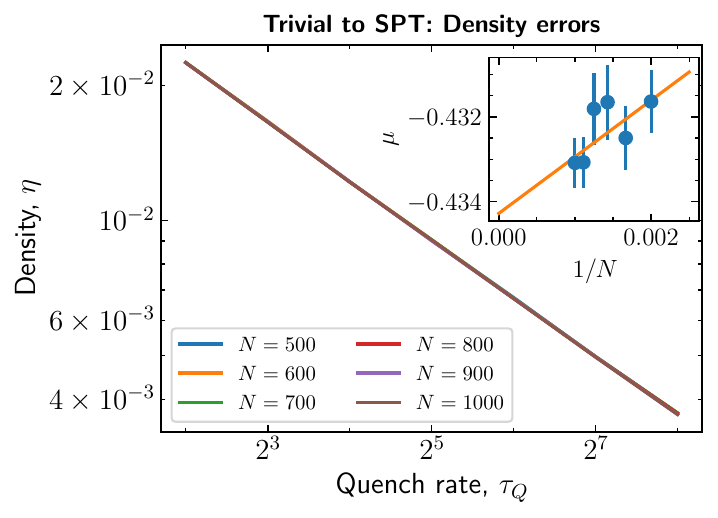}} 
    &&
    \subfig{(c)}{\includegraphics[width=.3\linewidth]{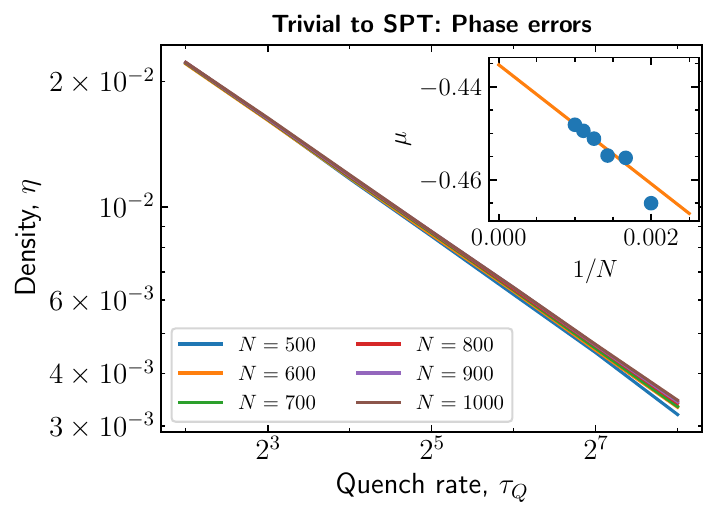}}
  \end{tabular}
\end{center}
\caption{Monte-Carlo sampling to obtain defect densities estimated by considering (a) all errors, (b) density 
errors, and (c) phase errors for the extended SSH model driven into the non-trivial SPT phase.
(Inset) Estimating the corresponding critical exponents by performing finite-size analysis leading
to $\mu=$ (a) 0.433(1), (b) 0.434(1), and (c) 0.435(2) that exhibit agreement with each other and also with the  
critical exponent obtained based on projector expectation values as in Sec.~\ref{sec:spt_order}.}  \label{figa:issh_d1pt0_mc}
\end{figure*}

As earlier, we drive the system into a non-trivial SPT phase and further sample the final evolved
wavefunction in the excitation basis. We compute the total defect density to be the sum of
density and phase fluctuations. Further, we determine the critical exponent as shown in
Fig.~\ref{figa:issh_d1pt0_mc}. The total defect densities can be approximated by considering either the 
defect densities or the phase densities. It is crucial to note that predictions of the QKZM still hold 
in the approximate limit of the total defect densities. In other words, it is possible to estimate the 
scaling exponent upto good accuracy based on partial defect density as illustrated in 
Fig.~\ref{figa:issh_d1pt0_mc}.

\section{Cross-validating the results by reversing the quench}
One other way to cross-validate the estimation of the scaling exponent is to reverse the quench direction.
For instance, in Sec.~\ref{sec:local_order}, we evaluated the scaling exponent by quenching a paramagnetic 
state to a ferromagnet. It should be possible to obtain the same scaling exponent by reversing the quench 
from a ferromagnet to a paramagnet as we drive through the same point of criticality as above. In this 
section, we validate the above notion by retrieving the scaling exponent for the phase transitions 
involving, both, local order and SPT order.

\begin{figure*}[h!]
\begin{center}
  \begin{tabular}{cp{0.01mm}cp{0.01mm}c}
    \subfig{(a)}{\includegraphics[width=.275\linewidth]{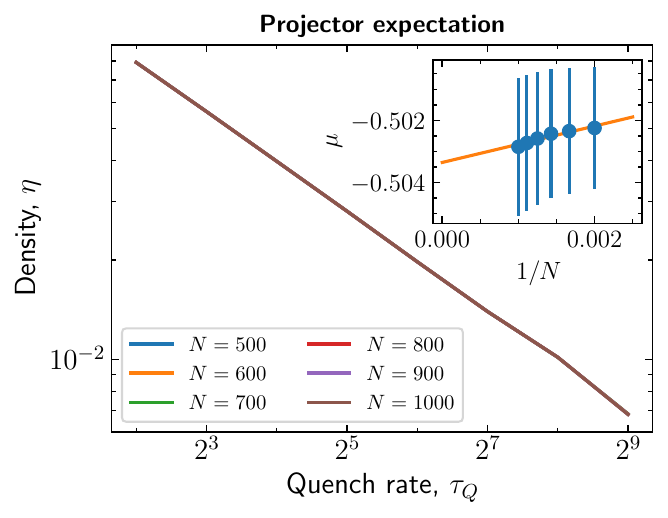}}
    &&
    \subfig{(b)}{\includegraphics[width=.3\linewidth]{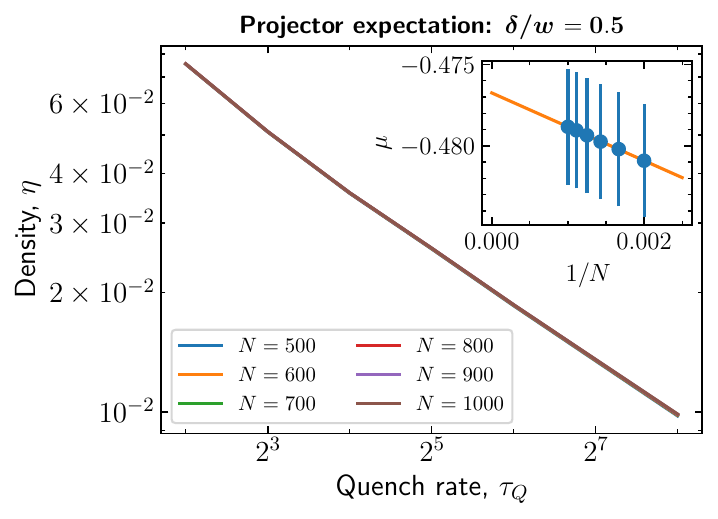}} 
    &&
    \subfig{(c)}{\includegraphics[width=.275\linewidth]{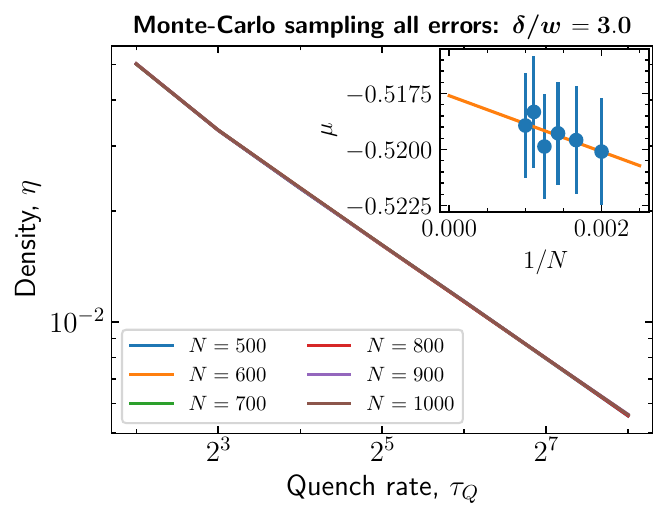}}
  \end{tabular}
\end{center}
\caption{Defect densities as a function of the quench rates for (a) Ising model, (b), (c) extended SSH model with 
$\delta/w$=0.5 and $\delta/w=3.0$ by reversing the direction of the quench. In (a) and (b) we compute the defect densities 
using the expectation value of the appropriate projector while in (c) we employ the Monte-Carlo sampling in the error 
basis. (Inset) Estimating the corresponding scaling exponent, $\mu$ by performing finite-size scaling analysis 
(a) 0.503(1), (b) 0.4767(3), and (c) 0.518(1) which are in good agreement with the corresponding scaling exponents obtained in 
the main text as in Fig.~\ref{fig:tfim}, Figs.~\ref{fig:essh} (b) and (d) respectively.}
\label{figb:reverse}
\end{figure*}

We begin by considering the case of quenching a ferromagnet to a paramagnet. In the protocol introduced in the main 
text the key to computing the defect density is the notion of errors with respect to a reference state. As we are
driving into a paramagnetic phase, we define the errors with respect to the reference state given by the paramagnetic groundstate $\ket{+}^{\otimes N}$, where $\ket{+} = \frac{1}{\sqrt{2}}(\ket{0} + \ket{1})$. Therefore, the total defect 
density can be estimated by defining $\mathcal{P}_{0}^{i} = \ket{+}\bra{+}_{i}$ in Eq.~\ref{eq:doe_exp}, leading 
to the estimation of the scaling exponent that is in good agreement with one obtained earlier in the main text, see 
Fig.~\ref{figb:reverse}(a). We note that the quench protocol adapted for this particular case is as in 
Eq.~\ref{eq:h_tdhahb} with $H_{I}$ being mapped to $-\sum\limits_{i=1}^{N-1}\sigma_{z}^{i}\sigma_{z}^{i+1}$ and 
$H_{F}$ being mapped to $-\sum\limits_{j=1}^{N}\sigma_{x}^{j}$.

We also cross-validate the scaling exponent in the context of SPT phases by evolving in the reverse direction 
i.e., from an initial state in the non-trivial SPT phase to a state in the trivial SPT phase. We consider 
the extended SSH Hamiltonian as in Eq.~\ref{eq:h_issh} and employ the quench protocol as in Eq.~\ref{eq1}.
As we are driving into the trivial SPT phase, we compute the defect density with respect to the reference 
state given by 
\begin{equation*}\label{eq:ssh_ham_gs_reverse}
 \ket{\phi}_{\text{AB}} = \frac{1}{\sqrt{2}}\prod_{i\in A}\left(\ket{0}_{i}\ket{1}_{i+1} - \ket{1}_{i}\ket{0}_{i+1}\right),
\end{equation*}
resulting in the total defect density given by 
\begin{equation}
\eta = \frac{1}{N}\sum_{i\in A}\left(1-\braket{\ket{\pmb{\minus}}\bra{\pmb{\minus}}}_{(i,i+1)}\right),
\label{eq:den_sshni_reverse}
\end{equation}
We retrieve the scaling exponent for the reverse quench and note that is in good agreement with the results as in 
Fig.~\ref{fig:essh} by computing the projector expectation for the case of $\delta/w=0.5$ and Monte-Carlo 
sampling in the error basis for $\delta/w=3.0$, see Figs.~\ref{figb:reverse}(b) and (c) respectively.

\section{Dependence on system size}
\begin{figure*}[h!]
\begin{center}
  \begin{tabular}{cp{0.01mm}cp{0.01mm}c}
    \subfig{(a)}{\includegraphics[width=.29\linewidth]{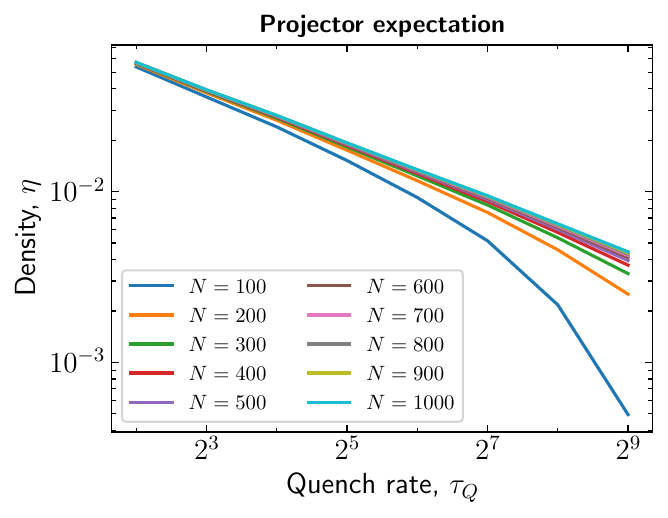}}
    &&
    \subfig{(b)}{\includegraphics[width=.29\linewidth]{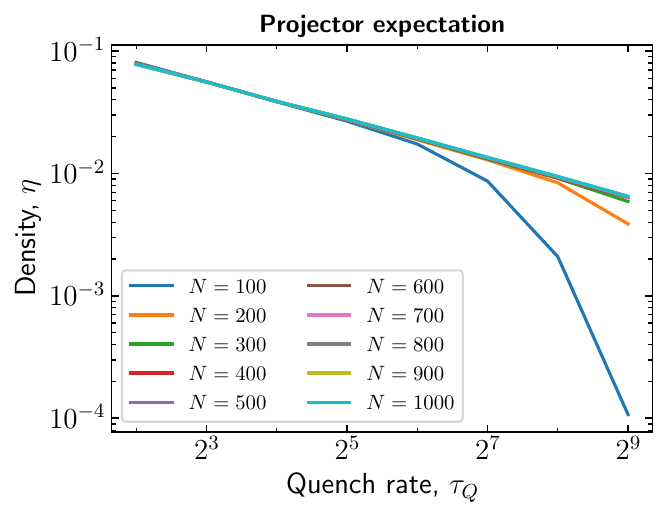}} 
    &&
    \subfig{(c)}{\includegraphics[width=.29\linewidth]{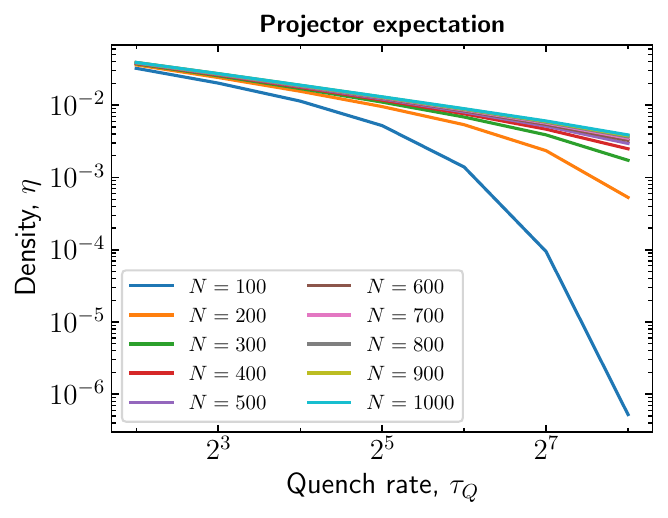}}
  \end{tabular}
\end{center}
\caption{Defect densities, $\eta$ as a function of the quench rates, $\tau_{Q}$ for (a) TFIM, (b) SSH and 
(c) ZXZ model. Across all the three models we note considerable finite-size scaling effects for system sizes $N<300$, 
however beyond the above limit the scaling of the defects follows the power-law (linear on $\log$-scale) prediction 
of the QKZM.}
\label{figc:system_size}
\end{figure*}

As the method described in the main text used to estimate the critical exponent, $\mu$ relies on 
finite-size scaling analysis, in this section, we present defect densities as a 
function of quench rates for different system sizes. We compute the above for the three models 
discussed in the main text i.e., the TFIM , SSH and the cluster state models as in Fig.~\ref{figc:system_size}. 
We adapt the quench protocol as in  Eq.~\ref{eq1} for all the three models while estimating the 
defect densities by computing the projector expectation value. We reiterate the use of DMRG for 
computing the initial state and TEBD for time evolution from the ITensor library.

\section{Scaling of defects as a function of time}
\begin{figure*}[h!]
\begin{center}
  \begin{tabular}{cp{0.01mm}cp{0.01mm}c}
    \subfig{(a)}{\includegraphics[width=.29\linewidth]{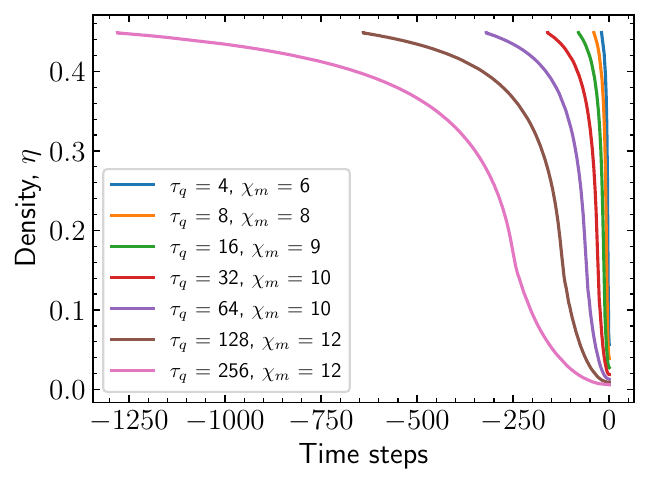}}
    &&
    \subfig{(b)}{\includegraphics[width=.29\linewidth]{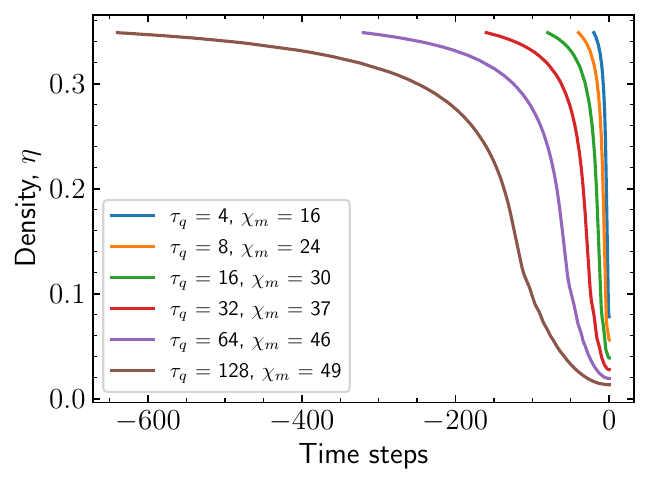}} 
    &&
    \subfig{(c)}{\includegraphics[width=.29\linewidth]{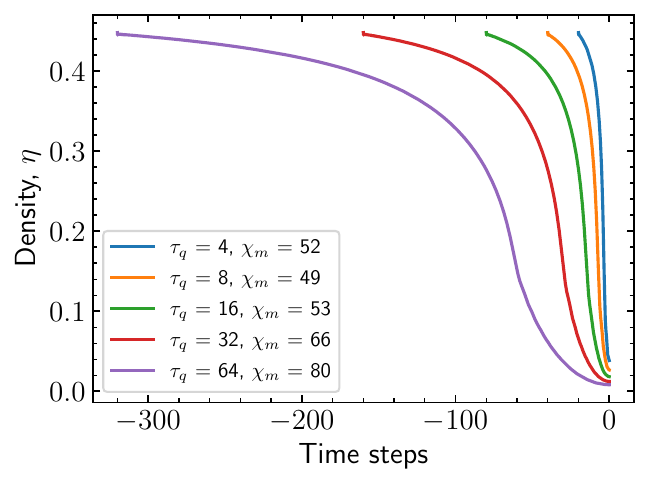}}
  \end{tabular}
\end{center}
\caption{Density of defects, $\eta$, as a function of the time steps with the maximum bond dimension, $\chi_{m}$ in the 
total evolution with error fixed to $10^{-8}$ for (a) TFIM, (b) SSH and (c) cluster state model for a fixed system size 
of $N=500$. We note that we fix the job run time on the cluster to be the same for the three models. The accessible
time steps (varying with $\delta t=0.05$) in the quench rates, $\tau_{Q}$ for the (a) TFIM in the same job run time is higher compared to (b) SSH further 
is higher compared to (c) cluster state model. We attribute this to the fact that the projector operator is two site and diagonal 
in the context of TFIM while is two site off-diagonal in the SSH model and is three site off diagonal in the context 
of cluster state model limiting the computation of expectation value for time steps of longer quench rates.}
\label{fige:den_bd_dt}
\end{figure*}

In this section, we present the defect density as a function of time by employing the quench protocol
as in Eq.~\ref{eq1} for a fixed system size of $N=500$. We compute the initial state using the DMRG protocol
of the ITensor library. Further, we compute the defect density at each time-step 
by evaluating the expectation value of the appropriate projection operator outlined in the main text for 
the three models of TFIM, SSH and the cluster state, while also recording the maximum bond dimension,
$\chi_{m}$ for each of the quench rates, $\tau_{Q}$, see Fig.~\ref{fige:den_bd_dt}.

\section{Performance metrics - Exact expectation, Monte-Carlo}
\begin{figure*}[h!]
\begin{center}
\includegraphics[width=0.875\linewidth]{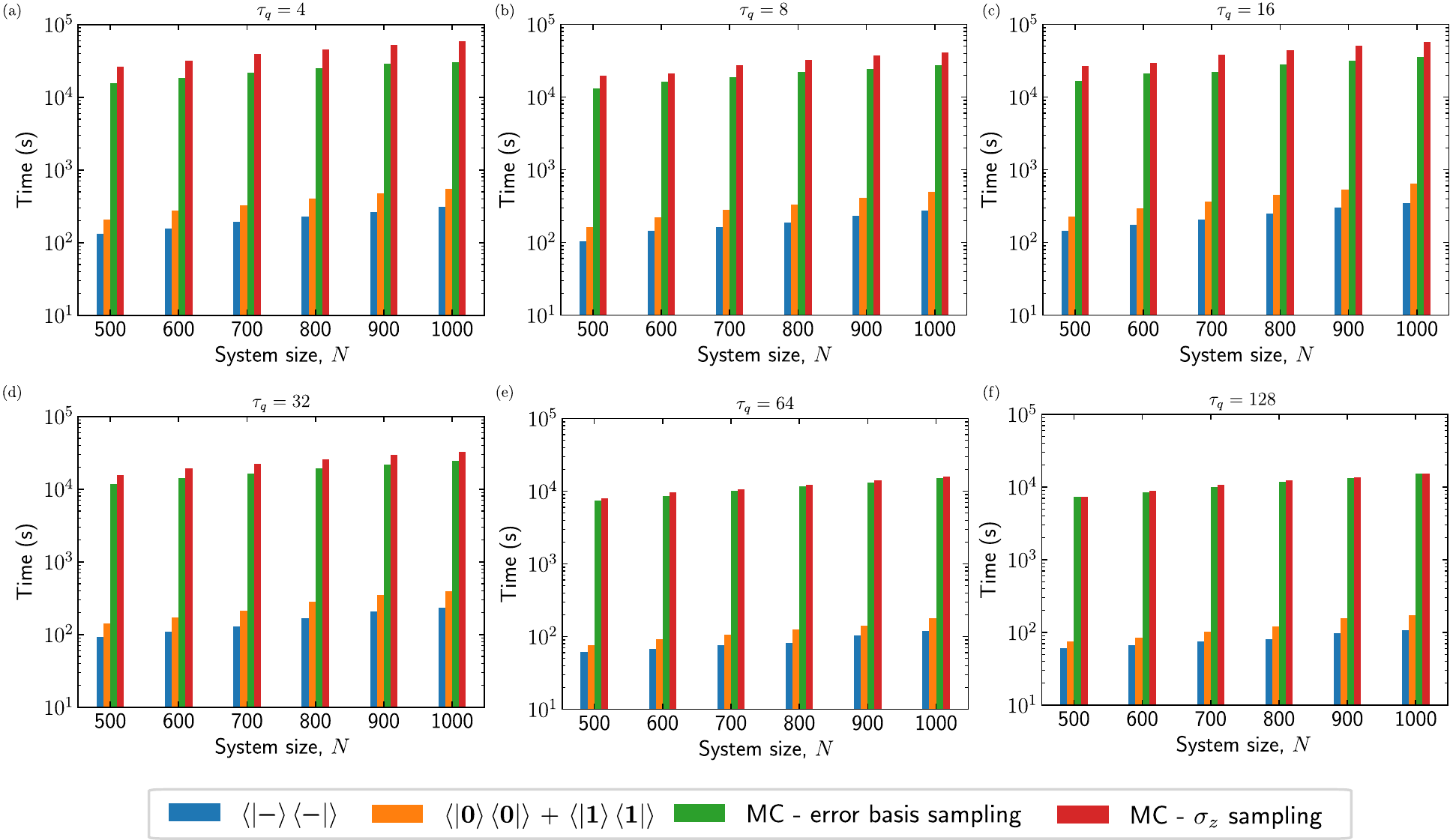}
\caption{Performance profiling of various routines used to compute the defect density of the extended 
SSH model as a function of the system size, $N$ for various fixed $\tau_{Q}$. We note that for fairness 
of comparison all the performance runs are done using a single core. We observe that for a given $\tau_{Q}$ 
the computation of defect density using the expectation value of two-site off-diagonal projector is faster 
compared to the expectation value of two-site diagonal projection operator. The performance of the latter
can be further optimized by considering the clubbed two-site diagonal projection operator (in the current 
scenario, we report the unclubbed version that involves evaluating the expectation value of two diagonal 
operators independently). We note that the Monte-Carlo sampling methods are considerably slower and also can
gain in performance by using parallelization techniques. In comparing both the Monte-Carlo techniques, the two-site
sampling is faster in comparison to the single-site sampling as the effective system size is halved in the former
leading to gain in performance. We also note that time to solution for computing defect density decreases with 
increase in $\tau_{Q}$. In other words, computing projector expectation value/sampling the evolved state
is computationally easier at longer quench rates compared to shorter quench rates, $\tau_{Q}$.}
\label{figd1:perf_tq_fixed}
\end{center}
\end{figure*}

\begin{figure*}[h!]
\begin{center}
\includegraphics[width=0.875\linewidth]{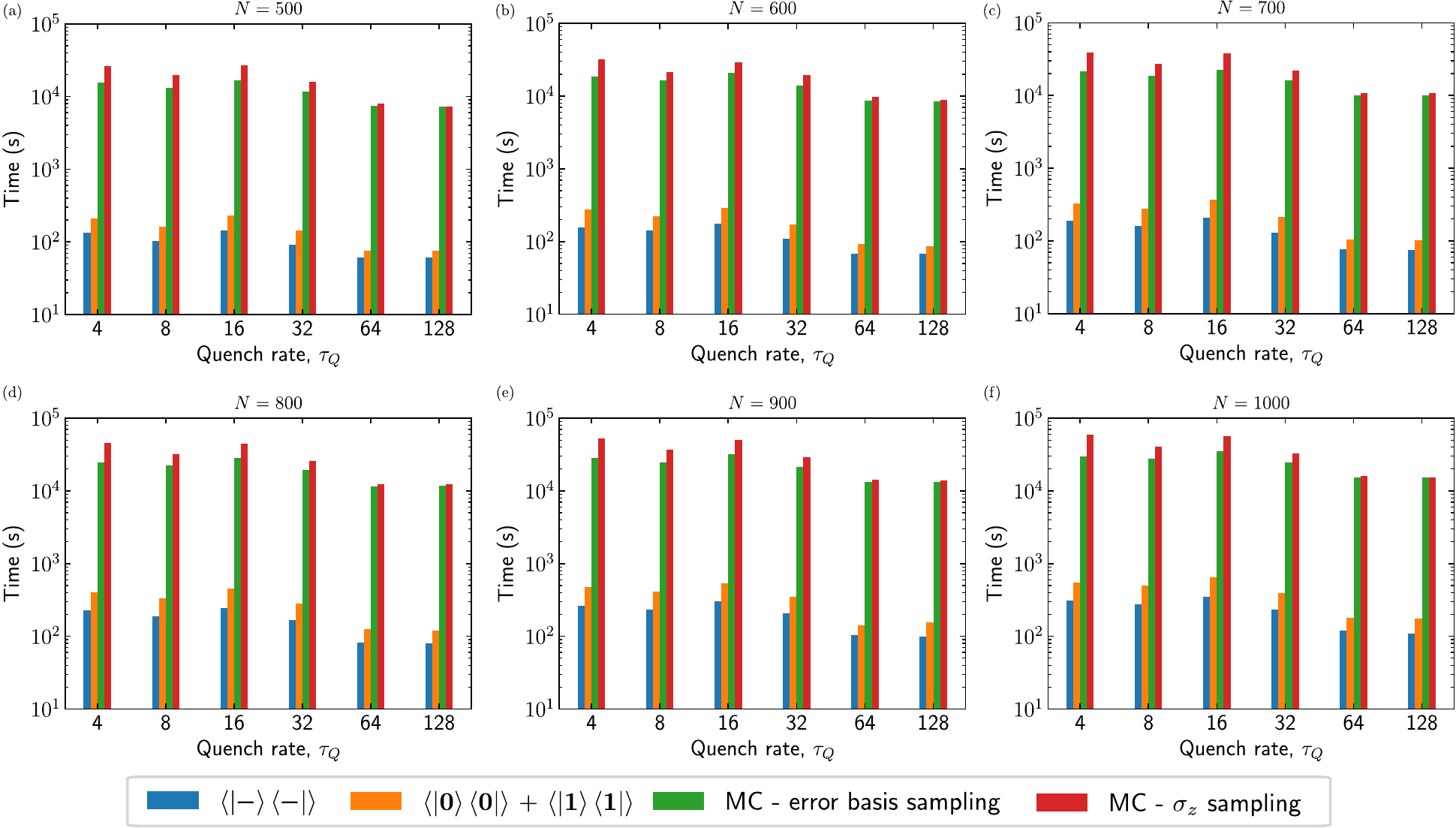}
\caption{Profiling the performance of defect density computation as a function of the quench rate, $\tau_{Q}$
for a given system size, $N$. In contrast to the above, it is tough to observe generic trends except for the 
fact that across all system sizes the time to compute the projector expectation value/sampling the errors is 
smaller at longer quench rates in comparison to shorter quench rates that demand longer computational time.}
\label{figd2:perf_N_fixed}
\end{center}
\end{figure*}

In this section, we report the performance of different methods deployed in computing the defect density. 
We consider the case of the extended SSH model and the quench protocol as on a digital quantum computer as in 
Eq.~\ref{eq:h_sshi}. For the rest of the analysis we further consider the case of $\delta/w=3$ and report the time 
to solution of computing the defect densities using the Julia macro $@elapsed$. We profile the following four 
different strategies that compute the defect density (a) Expectation value of the
projector operator $\ket{\pmb{-}}\bra{\pmb{-}}$, (b) Expectation value of the density fluctuations, (c) Monte-Carlo
sampling in the full-error basis and (d) Monte-Carlo sampling in the $\sigma_{z}$ basis and report 
the profiling as a function of the system size, $N$ for different quench rates, $\tau_{Q}$ as in 
Fig.~\ref{figd1:perf_tq_fixed}, while report the profiling as a function of the quench rate, $\tau_{Q}$
for different system sizes, $N$, see Fig.~\ref{figd2:perf_N_fixed}. For fairness of comparison, we benchmark
all the runs using a singlethread. However, we note that the Monte-Carlo simulation can be made run faster
by utilizing parallelization routines (multi-node multi-core spread using the macro @$parallel$).

\section{Statistics of defects and scaling of cumulants}
In Ref.~\onlinecite{del_Campo_2018}, the authors establish that the statistics of the defect number corresponding 
to the final evolved state obtained by quenching across a quantum phase transition respect the power law scaling 
as in the QKZM. In this section, we extend the above to the context of quantum phase transitions with SPT 
order while also reproducing the results in the context of quantum phase transitions with local order.

\begin{figure*}[h!]
\begin{center}
  \begin{tabular}{cp{0.01mm}cp{0.01mm}c}
    \subfig{(a)}{\includegraphics[width=.275\linewidth]{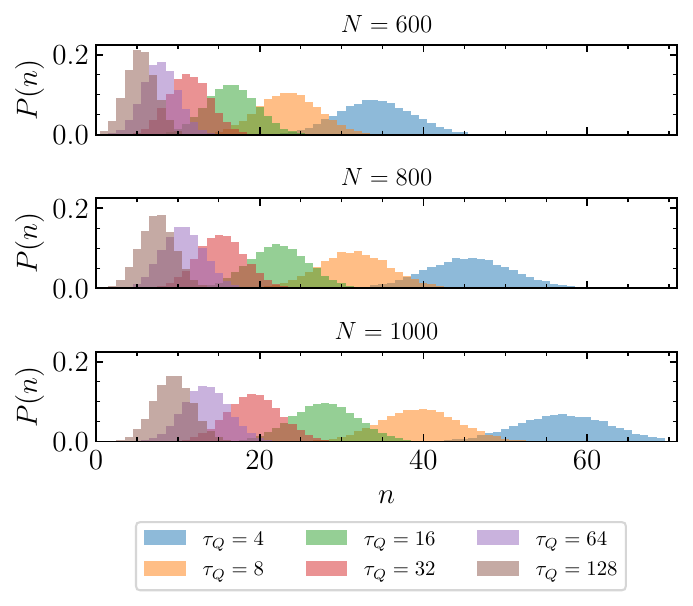}}
    &&
    \subfig{(b)}{\includegraphics[width=.3\linewidth]{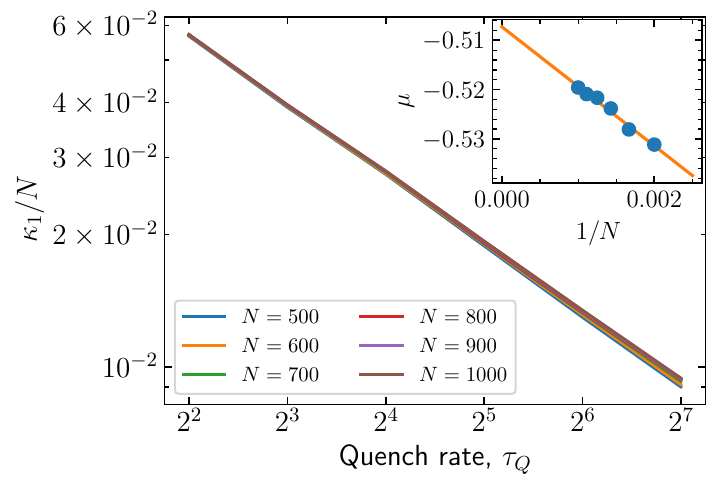}} 
    &&
    \subfig{(c)}{\includegraphics[width=.3\linewidth]{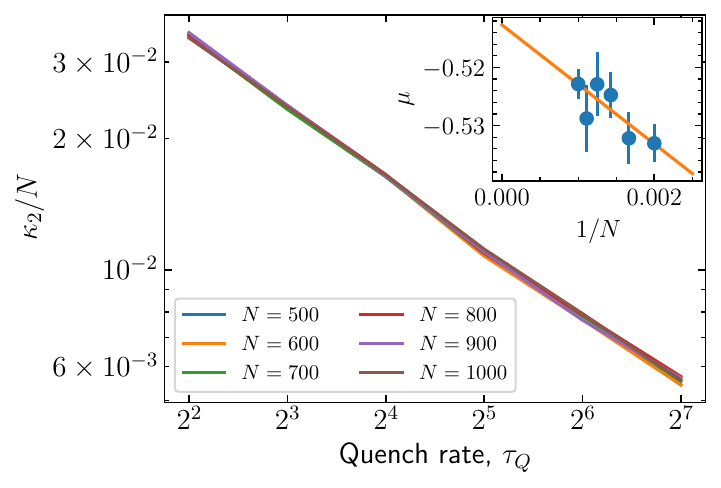}}
  \end{tabular}
\end{center}
\caption{Defect number probability distribution and its related properties in the case of TFIM dynamics. 
(a) Defect number distribution for different quench rates $\tau_{Q}$ for a given system size (top) $N=600$, 
(middle) $N=800$, (bottom) $N=1000$. (b) first cumulant, $\kappa_1/N$ (c) second cumulant, $\kappa_2/N$, both 
scale as power-law of the quench rate, $\tau_{Q}$ with the scaling exponent obtained by performing finite-size
scaling analysis, $\mu=$ (b)  0.507(1) (c) 0.513(4).
\label{figf:ising}}
\end{figure*}

To this extent, we observe that the Monte-Carlo based protocol used to estimate the defect density 
can also be extended to study the statistics of the number of defects. That is, the Monte-Carlo sampling can 
be used to generate error strings, further leading to the construction of the defect/error number distribution.
We begin by considering the TFIM model as in Eq.~\ref{eq:h_tfim} and quench a paramagnetic initial state 
to a final state belonging to the ferromagnetic phase. We employ the Monte-Carlo based sampling
as introduced in the main text to construct the probability distribution of the number of defects i.e., 
for each trajectory we sample over the $N$ sites in the $\sigma_{z}$ basis and compute the number of 
defects and repeat for many trajectories, resulting in the probability distribution as in 
Fig.~\ref{figf:ising}(a). We note that the probability distribution of the number of defects is qualitatively in 
agreement with the results in Ref.~\onlinecite{del_Campo_2018}. Further, we compute the first and 
second cumulants (mean and variance) of the distribution and note that they exhibit a power law 
scaling with the quench rate, see Fig.~\ref{figf:ising}(b, c). In a similar fashion it is possible to compute
the scaling of higher cumulants and we leave this exercise to the future as more data points at intermediate 
quench rates are required to accurately capture the scaling.

\begin{figure*}[h!]
\begin{center}
  \begin{tabular}{cp{0.01mm}cp{0.01mm}c}
    \subfig{(a)}{\includegraphics[width=.275\linewidth]{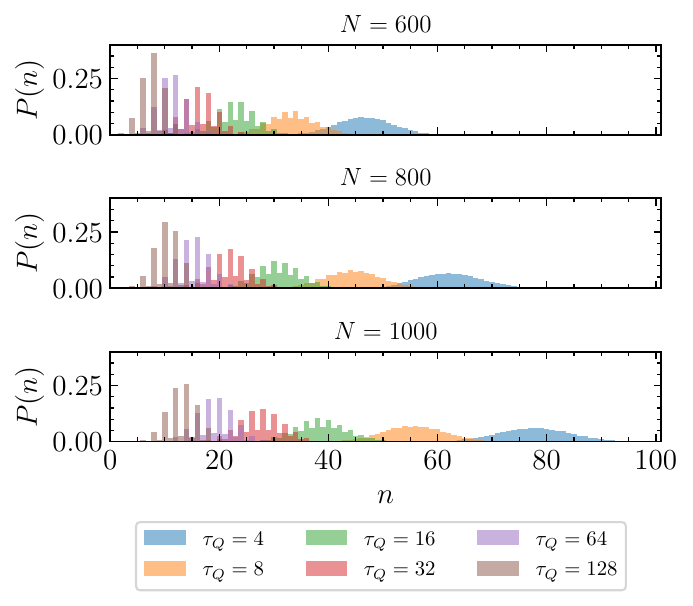}}
    &&
    \subfig{(b)}{\includegraphics[width=.3\linewidth]{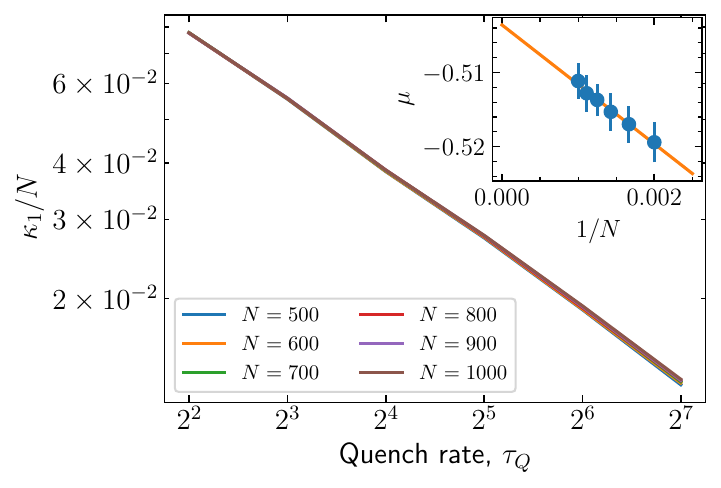}} 
    &&
    \subfig{(c)}{\includegraphics[width=.3\linewidth]{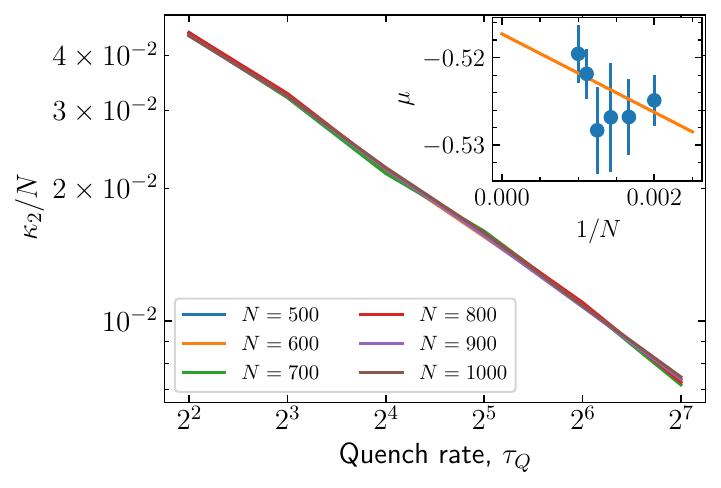}}
  \end{tabular}
\end{center}
\caption{Probability distribution of number of defects and higher order cumulants in the case of SSH dynamics.
(a) Defect number distribution for different quench rates $\tau_{Q}$ for a given system size (top) $N=600$, 
(middle) $N=800$, (bottom) $N=1000$. We note the distribution is dominated by even number of defects, a potential
reason being the equal number of density fluctuations, a validation of which we leave for future exploration.
We observe that the first and the second cumulant, (b) $\kappa_{1}/N$ (c) $\kappa_{2}/N$ exhibit power-law
scaling with $\mu=$ (b) 0.504(1), (c) 0.517(4) as in the QKZM.}  
\label{figf:ssh}
\end{figure*}

In a similar fashion, we extend the analysis to the quench protocol of the SSH model as in Eq.~\ref{eq:td_ssh_ham}
wherein we quench a state belonging to the trivial SPT phase to a state belonging to the non-trivial SPT phase.
We deploy the Monte-Carlo sampling procedure and sample the final evolved state in the error basis given by 
$\ket{\pmb{0}}$, $\ket{\pmb{1}}$, $\ket{\pmb{\plus}}$. We further construct the probability distribution 
of the number of defects as in Fig.~\ref{figf:ssh}(a) and further compute the higher order cumulants, see
Fig.~\ref{figf:ssh}(b), (c) that exhibit a power-law scaling as in the previous scenario.

\begin{figure*}[h!]
\begin{center}
  \begin{tabular}{cp{0.01mm}cp{0.01mm}c}
    \subfig{(a)}{\includegraphics[width=.275\linewidth]{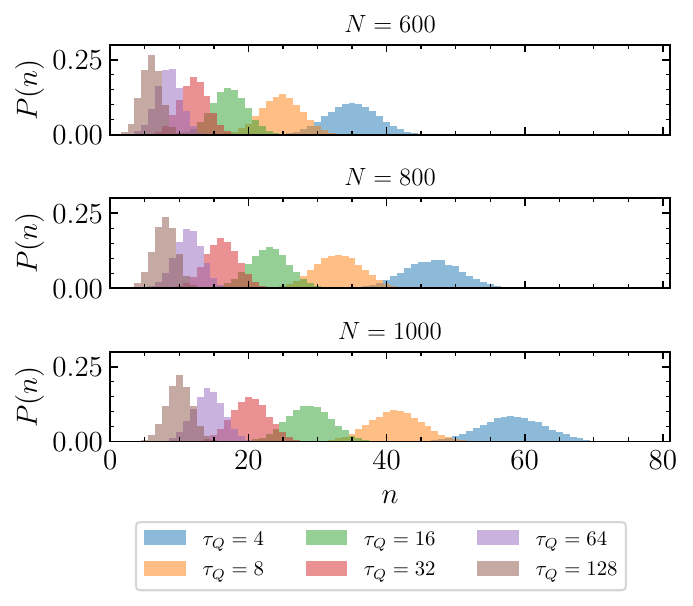}}
    &&
    \subfig{(b)}{\includegraphics[width=.3\linewidth]{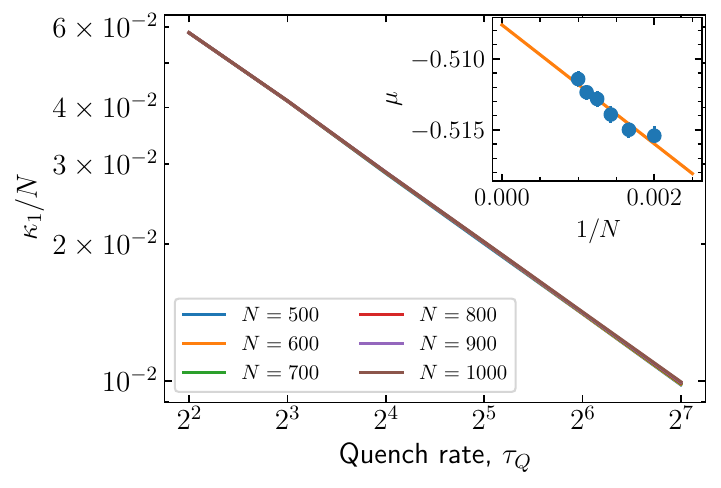}} 
    &&
    \subfig{(c)}{\includegraphics[width=.3\linewidth]{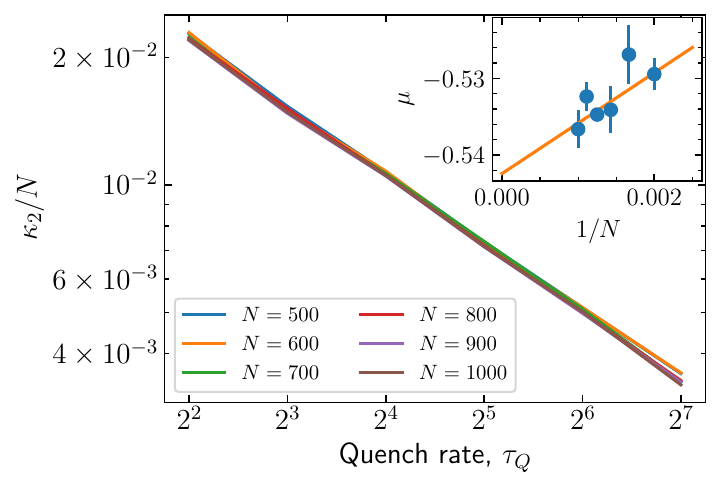}}
  \end{tabular}
\end{center}
\caption{Defect number probability distribution and the scaling of its cumulants for the case of the dynamics of 
the extended SSH model with $\delta/w=3$. We sample the defects in the error basis given by $\{\ket{\pmb{0}}, \ket{\pmb{1}}, 
\ket{\pmb{\plus}}\}$. (a) Defect number distribution for different quench rates, $\tau_{Q}$ for a given system 
size (top) $N=600$, (middle) $N=800$, (bottom) $N=1000$. As opposed to the previous case where the distribution
was dominated by even number of errors, here we observe a more continuous distribution. The scaling of the 
(b) first cumulant, $\kappa_{1}/N$ and (c) second cumulant, $\kappa_{2}/N$ both exhibiting a power-law 
scaling with $\mu=$ (b) 0.508(1), (c) 0.542(4).}
\label{figf:issh_minus}
\end{figure*}

\begin{figure*}[h!]
\begin{center}
  \begin{tabular}{cp{0.01mm}cp{0.01mm}c}
    \subfig{(a)}{\includegraphics[width=.275\linewidth]{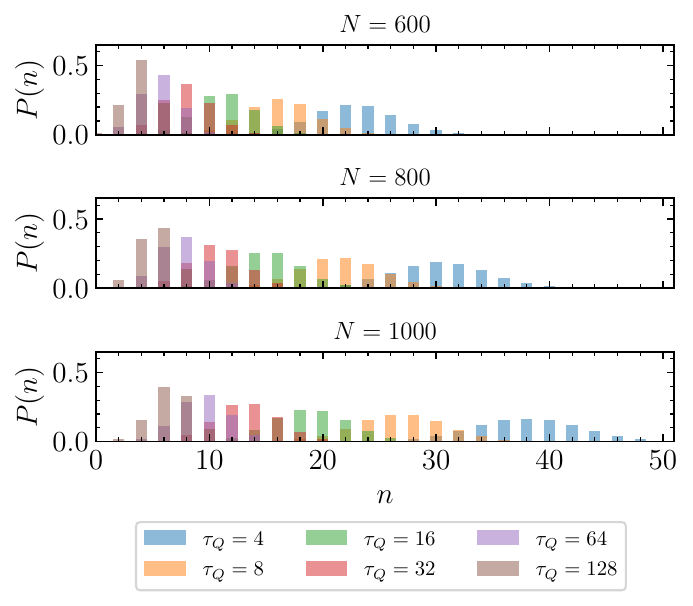}}
    &&
    \subfig{(b)}{\includegraphics[width=.3\linewidth]{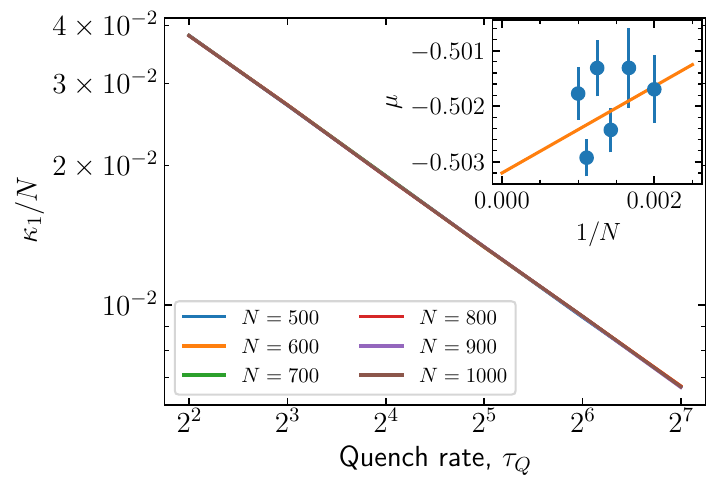}} 
    &&
    \subfig{(c)}{\includegraphics[width=.3\linewidth]{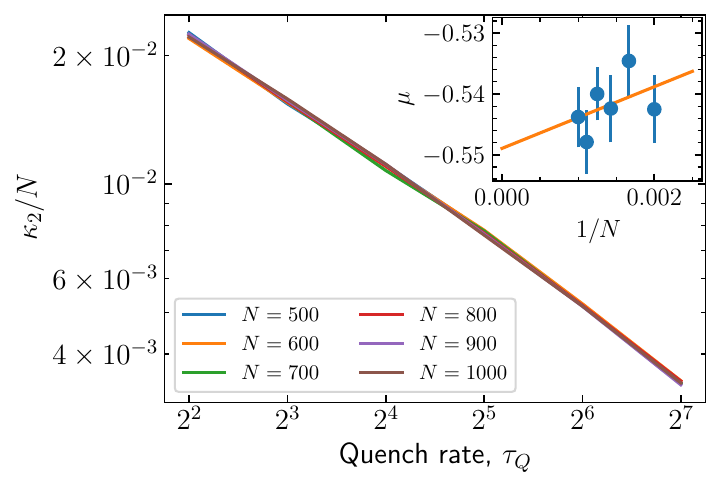}}
  \end{tabular}
\end{center}
\caption{Defect number probability distribution for the case of the dynamics of the extended SSH model as on 
a digital quantum computer. We sample the defects in the truncated basis i.e., $\sigma_{z}$ basis and report the
(a) probability distribution of the number of defects, (b) first cumulant, $\kappa_{1}$/N and (c) second
cumulant, $\kappa_{2}/N$ both retrieving the QKZM scaling with $\mu=$ (b) 0.503(1), (c) 0.549(7).}  
\label{figf:issh_zo}
\end{figure*}

We further extend the analysis to the case of the dynamics of the extended SSH model as in 
Eq.~\ref{eq:h_sshi} that allows for the realization of the dynamics on a digital quantum computer i.e., we 
evolve across a tivial to non-trivial topological phase transition with $\delta/w=3$. As earlier, we sample 
the final evolved wavefunction in the error basis to construct the probability distribution of the number of 
defects, see Fig.~\ref{figf:issh_minus}(a) and further obtain the scaling properties of its cumulants, see 
Fig.~\ref{figf:issh_minus}(b, c) that exhibit a power-law scaling with respect to the quench rate, $\tau_{Q}$
as previously. The truncated sampling in the $\sigma_{z}$ basis instead of the entire error basis allows for
the study of the probability distribution of number of defects (density defects) and the scaling properties
of its cumulants, see Fig.~\ref{figf:issh_zo}. We note that even in the case of the truncated sampling 
in $\sigma_{z}$ basis we observe that the QKZM scaling is respected. However, in both the above scenarios 
we notice that scaling exponent of the second cumulant deviates from the value of 0.5 by a considerable 
amount yet respects the power law scaling.

\section{Inhomogenous Ising model}
\begin{figure*}
\begin{center}
  \begin{tabular}{cp{0.01mm}cp{0.01mm}c}
    \subfig{(a)}{\includegraphics[width=.3\linewidth]{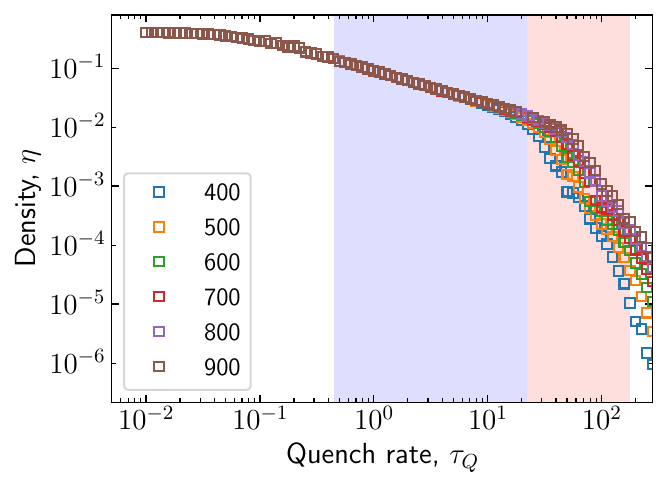}}
    &&
    \subfig{(b)}{\includegraphics[width=.3\linewidth]{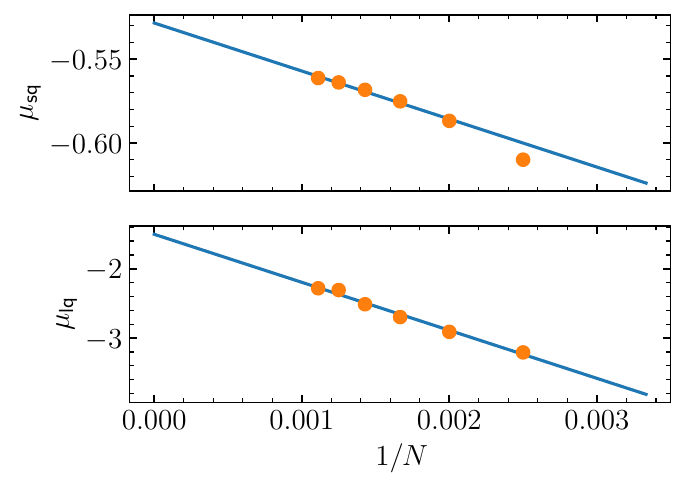}} 
    &&
    \subfig{(c)}{\includegraphics[width=.3\linewidth]{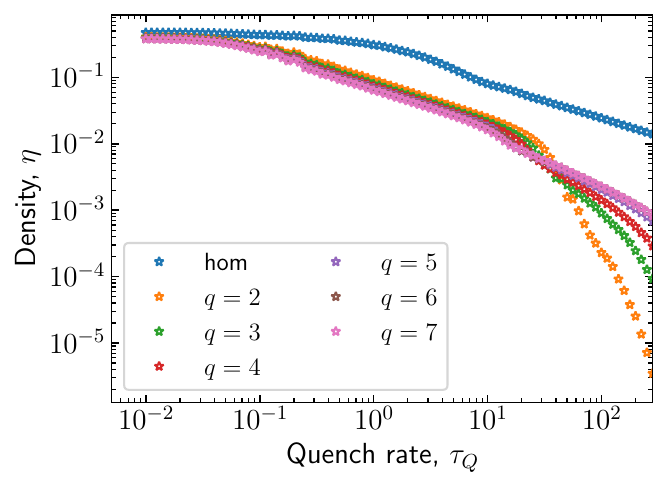}}
  \end{tabular}
\end{center}
\caption{Estimating the scaling exponent in the inhomogenous TFIM. (a) Defect density, $\eta$ as a function
of the quench rate, $\tau_{Q}$. The regions marked in (blue) red are used to estimate the scaling exponent 
for (short) long  quench rates for a given system size. (b) Finite-size scaling analysis to obtain the 
scaling exponent in the thermodynamic limit for (top) shorter quench rates, $\mu_{\text{sq}}=$0.528(4)  
(bottom) longer quench rates, $\mu_{\text{lq}}=$1.50(6) which are in good agreement with the values 
obtained in Ref.~\onlinecite{G_mez_Ruiz_2019}. (c) Defect density, $\eta$ as a function of the quench rates, 
$\tau_{Q}$ with increasing $q$ as in the interaction function, given by $J_{q}(n)$ as in Eq.~\ref{eqg:jqn}. 
We note that with increase in $q$ the behavior of the defect density approaches the homogenous case where 
$J_{q}(n)=J$, a constant. We also note that the qualitative behavior of the defect density remains similar 
as in the above reference.}
\label{figg:ising_inhom}
\end{figure*}

In Ref.~\onlinecite{G_mez_Ruiz_2019}, the authors study the QKZM in an inhomogenous setup. They estimate
the scaling exponent for short and long quench rates, denoted by $\mu_{\text{sq}}$ and $\mu_{\text{lq}}$ 
respectively, in the TFIM. The Hamiltonian governing the dynamics is 
given by
\begin{equation*}
H_{\text{IFTIM}} = -\sum\limits_{i=1}^{L-1}J_{q}(n)\sigma_{z}^{i}\sigma_{z}^{i+1} - h(t)\sum\limits_{j=1}^{L}\sigma_{x}^{j}
\end{equation*}
where the nearest neighbor interaction is modulated by $J_{q}(n)$,
\begin{equation*}
J_{q}(n) = J(0)(1-\alpha|n|^{q}),
\label{eqg:jqn}
\end{equation*}
where $n$ are the sites of nearest neighbor interaction, $J(0)$ being the interaction at the end of the chain,
$\alpha$ is a constant such $J(0)=J$ recovers the homogeneous interaction at the ends of the open chain. For a more
detailed description of $J_{q}(n)$, we refer the interested reader to Fig.~1 in Ref.~\onlinecite{G_mez_Ruiz_2019}.
We consider the quench dynamics given by
\begin{equation}
h(t) = J(0)\bigg(1-\frac{t}{\tau_{Q}}\bigg), t \in [-\tau_{Q}, \tau_{Q}],
\end{equation}
i.e., we quench a initial state in the paramagnetic phase to a state in the ferromagnetic phase.
In Fig.~\ref{figg:ising_inhom}, we study the QKZM for short and long quench rates and further estimate 
the scaling exponent by computing the defect density using the projector expectation value as in 
Eq.~\ref{eq:exp_tfim} in both regimes for the case of $q=2$. We note that the results obtained using
our method are in good agreement with the results obtained in the Ref.~\onlinecite{G_mez_Ruiz_2019}. 
Further, for a fixed system size of $N=500$, in Fig.~\ref{figg:ising_inhom}(c), we also present the scaling 
of defect density, computed using Eq.~\ref{eq:exp_tfim} for short and long quench rates for increasing $q$ 
whose behavior is qualitatively in agreement with the results in the above reference. We note that the 
critical exponents estimated in the current scenario are sensitive to the finite-size scaling analysis, and we postpone this and related analysis on the critical exponents for various $q$ to a future work.

\section{Extensions to systems in higher dimensions\label{app:h}}
In this section, we briefly sketch an outline for estimating the scaling exponent of systems in higher dimensions
by considering two paradigmatic models: the TFIM with local order and the toric code with topological 
order~\cite{Kitaev_2003}, both in two dimensions. Errors being central to the above estimation, we will introduce 
the errors for the above models and leave the numerical estimation to the future.

In Ref.~\onlinecite{Schmitt_2022}, the authors have not only studied the phase diagram of the 
2-dimensional (2D) TFIM with ferromagnetic interaction but also estimate the scaling exponent by 
considering the perturbed dynamics. With reference to the above model, the errors can be estimated using the 
Eq.~\ref{eq:exp_tfim} by considering all the nearest neighbor interactions. We also note that the dynamics 
of the 2D TFIM can be simulated by techniques as listed in the above reference.

One of the well-known models exhibiting topological order in 2D is the toric code model. To describe the 
perturbation free Hamiltonian of the toric code, we consider a 2D square lattice with spins on the edges of the 
lattice. Next, we introduce the vertex/face operator, $A_{v}/B_{p}$ given by $\prod_{i}\sigma_{x}^{i}$/
$\prod_{i}\sigma_{z}^{j}$ where $i$/$j$ are the spins attached to the vertices and the faces respectively. The 
Hamiltonian is thus given by $H = -\sum_{v}A_{v} - \sum_{p}B_{p}$. Recent works, for instance 
Refs.~\onlinecite{Jamadagni2022a, Jamadagni2023} have characterized topological order using errors of the toric 
code, given by the violations of the $A_{v}$ and $B_{p}$ operators i.e., $(\mathds{1} - A_{v})/2$ and $(\mathds{1} - B_{p})/2$. Therefore, by considering a time-dependent perturbation that drives a trivial 
state into a non-trivial topological state and estimating the defect density using the above projection 
operators provides a pathway for the estimation of the scaling exponent.



\end{document}